\documentclass[preprint2,trackchanges]{aastex61}
\pdfoutput=1 %for arXiv submission
\usepackage{amsmath,amstext}
\usepackage[T1]{fontenc}
\usepackage{apjfonts} 
\usepackage[figure,figure*]{hypcap}

\newcommand{\Mearth}{M_{\oplus}}

\newcommand{\fracbrac}[2]{\left(\frac{#1}{#2}\right)}

\newcommand{\gsim}{\lower.7ex\hbox{$\;\stackrel{\textstyle>}{\sim}\;$}}
\newcommand{\lsim}{\lower.7ex\hbox{$\;\stackrel{\textstyle<}{\sim}\;$}}

\newcommand{\BBa}{BB16a}
\newcommand{\BBb}{BB16b}
\newcommand{\au}{\text{AU}}

\newcommand{\PX}{Planet Nine}

\newcommand{\figref}[1]{Figure \ref{#1}}
\shorttitle{Chaotic Dynamics and Planet-9}
\shortauthors{Hadden et al.}

\submitjournal{AJ}
\begin{document}

%%% Title 
\title{Chaotic Dynamics of Trans-Neptunian Objects Perturbed by Planet Nine}

\author{Sam~Hadden}
\affiliation{Harvard-Smithsonian Center for Astrophysics, 60 Garden St., MS 51, Cambridge, MA 02138, USA}
\correspondingauthor{Sam~Hadden}
\email{samuel.hadden@cfa.harvard.edu}

\author{Gongjie~Li}
\affiliation{Harvard-Smithsonian Center for Astrophysics, 60 Garden St., MS 51, Cambridge, MA 02138, USA}
\affiliation{Center for Relativistic Astrophysics, School of Physics, Georgia Institute of Technology, Atlanta, GA 30332, USA}
\author[0000-0001-5133-6303]{Matthew~J.~Payne} 
\affiliation{Harvard-Smithsonian Center for Astrophysics, 60 Garden St., MS 51, Cambridge, MA 02138, USA}

\author[0000-0002-1139-4880]{Matthew~J.~Holman}
\affiliation{Harvard-Smithsonian Center for Astrophysics, 60 Garden St., MS 51, Cambridge, MA 02138, USA}

%%%%%%%%%%%%%%%%%%%%%%%%%%%%%%%%%%%%%%%%%%%%%%%%%%%%%%%%%%%%%%%%%%%%%%%%%

%%%%%%%%%%%%%%%%%%%%%%%%%%%%%%%%%%%%%%%%%%%%%%%%%%%%%%%%%%%%%%%%%%%%%%%%%
\begin{abstract}
Observations of clustering among the orbits of the most distant trans-Neptunian objects (TNOs) has inspired interest in the possibility of an undiscovered ninth planet lurking in the outskirts of the solar system.
Numerical simulations by a number of authors have demonstrated that, with appropriate choices of planet mass and orbit, such a planet can maintain clustering in the orbital elements of the population of distant TNOs, similar to the observed sample. 
However, many aspects of the rich underlying dynamical processes induced by such a distant eccentric perturber have not been fully explored.
We report the results of our investigation of the dynamics of coplanar test-particles that interact with a massive body on an  circular orbit (Neptune) and a massive body on a more distant, highly eccentric orbit (the putative \PX).
We find that a detailed examination of our idealized simulations affords tremendous insight into the rich test-particle dynamics that are possible.
In particular, we find that chaos and resonance overlap plays an important role in particles' dynamical evolution.
We develop a simple mapping model that allows us to understand, in detail, the web of overlapped mean-motion resonances explored by chaotically evolving particles.
We also demonstrate that gravitational interactions with Neptune can have profound effects on the orbital evolution of particles.
Our results serve as a starting point for a better understanding of the dynamical behavior observed in more complicated simulations that can be used to constrain the mass and orbit of \PX.
\end{abstract}
%%%%%%%%%%%%%%%%%%%%%%%%%%%%%%%%%%%%%%%%%%%%%%%%%%%%%%%%%%%%%%%%%%%%%%%%%

%%%%%%%%%%%%%%%%%%%%%%%%%%%%%%%%%%%%%%%%%%%%%%%%%%%%%%%%%%%%%%%%%%%%%%%%%
% Kuiper belt objects: individual (\NAME)
%%%%%%%%%%%%%%%%%%%%%%%%%%%%%%%%%%%%%%%%%%%%%%%%%%%%%%%%%%%%%%%%%%%%%%%%%

%%%%%%%%%%%%%%%%%%%%%%%%%%%%%%%%%%%%%%%%%%%%%%%%%%%%%%%%%%%%%%%%%%%%%%%%%
\section{Introduction}
\label{SECN:INTRO}
\citet{Trujillo.2014} noted that the then-known ``extreme scattered disk objects'' (ESDOs) with semimajor axes greater than $150\au$ and perihelion distances greater than $30\au$ had arguments of pericenter, $\omega$, clustered around $0^{\circ}$.
Subsequently, \citet{Batygin.2016} (\BBa) noted that the six dynamically stable, distant ESDOs that were known at the time were all apsidally and {\it nodally} aligned, having their long axes roughly aligned \emph{and} sharing the same orbital plane. 
Both \citet{Trujillo.2014} and \BBa~suggested that the observed clustering in orbital elements could be maintained by a distant, unseen planetary-mass perturber.

A number of authors had previously suggested that various aspects of the distribution of outer solar system objects could be explained by the presence of an additional planetary mass object \citep[e.g.][]{Brunini.2002,Melita.2004,Gladman.2006,Gomes.2006,Lykawka.2008}.
\BBa~promulgated a specific hypothesis: a Neptune-mass (~$10-15 M_\earth$) planet in a distant (~$a\sim 700$~\au), eccentric ($e\sim0.6$), and inclined ($i\sim30\arcdeg$) orbit is responsible for the orbital clustering seen among ESDOs.  
This planet would share the orbital plane of the ESDOs, but it would be apsidally anti-aligned to the (then-known) cluster of ESDOs.  

The astronomy community has been invigorated by this suggestion, leading to a wide range of related topics being investigated, including:
constraining its location \citep{Brown.2016, HP16a, HP16b}; 
formation mechanisms \citep{Bromley16,Kenyon16,Li16,Mustill16}, 
atmospheric signatures \citep{Fortney16,Levi17}, 
dynamics \citep{Malhotra.2016, Veras16,BatyginMorbidelli17, Becker2017,Millholland17}, 
influences on trans-Neptunian objects (TNOs) \citep{BB16c, SHANKMAN17a}, 
generation of solar obliquity \citep{Bailey16,LAI16,Gomes17}, 
and many more. 

\BBa~and \citet{Brown.2016} (hereafter \BBb) demonstrated that the \emph{result} of introducing a distant, eccentric planet into dynamical simulations of the outer solar system is the clustering, at late-times, of test-particle orbits into a configuration that is anti-aligned with the planet; however the \emph{mechanism} by which this anti-alignment is generated has remained rather unclear. 
Since then, \citet{BEUST16} and \citet{BatyginMorbidelli17} have pointed out that, in the secular approximation, libration islands appear in the phase space for high-eccentricity, anti-aligned configurations.  
This anti-aligned libration would explain how TNOs maintain their anti-alignment with \PX, to the degree that the secular approximation is valid.

The secular approximation breaks down whenever there are any resonant or short-term interactions, such as a close encounter.
Consequently, the usual secular approximation {\it must} break down for any anti-aligned ESDOs   orbiting in (nearly) the same plane as \PX: close encounters are an eventuality for such ESDOs, unless they are protected from encounters by the phase-protection of a mean motion resonance (MMR) with \PX.
Indeed, \citet{BatyginMorbidelli17} find that particles surviving the full 4 Gyr duration of their simplified coplanar $N$-body simulations of ESDOs perturbed by \PX~are able to do so through such phase protection.
\citet{BEUST16} and \citet{BatyginMorbidelli17} also demonstrate, using a modified secular averaging accounting for MMRs between ESDOs and \PX, that such resonant ESDOs can maintain permanent anti-alignment with \PX's orbit. \citet{BatyginMorbidelli17}  propose that this secular evolution in MMR is the fundamental dynamical mechanism responsible for producing the anti-aligned population observed in numerical simulations.
%and identified with the clustered population of observed ESDOs.

\citet{BatyginMorbidelli17} find that, after introducing modest inclinations in their $N$-body simulations, particles no longer reside permanently in MMRs, but undergo chaotic semimajor axis evolution, "hopping" from resonance to resonance.
In Section \ref{SECN:NUMERICAL}, we will see that fully accounting for Neptune's gravitational potential in $N$-body simulations, rather than approximating its effect as a quadrupole potential (as \citet{BatyginMorbidelli17} do), also leads to chaotic diffusion of TNOs' semimajor axes.
In fact, this `active' Neptune drives significantly more chaotic diffusion than seen in \citet{BatyginMorbidelli17}'s simulations with modest inclinations.
\BBa~noted that the behavior of the distant TNOs in their simulation was suggestive of marginally overlapped mean-motion resonances through which the orbits could diffuse while still being protected from the large-scale scattering and ejection. 
A central aim of our work is to better understand the nature of this web of overlapped resonances and its role in chaotic evolution of TNOs.
The vigorous chaotic semimajor axis diffusion that we observe in our numerical simulations calls into question secular evolution in MMR as the fundamental mechanism maintaining the anti-alignment of ESDOs because particles generally do not spend a significant amount of time in any MMR.  
Therefore, we also revisit the dynamical mechanisms determining apsidal evolution of TNOs in the presence of \PX.

Resonance overlap is the fundamental mechanism by which chaos arises in most energy-conserving dynamical systems \citep[e.g.,][]{LLBook}.
%when applied to orbital dynamics has a rich history.
\citet{Chirikov79} \citep[and also][]{WalkerFord69} first proposed the heuristic `resonance overlap criterion' for predicting the onset of chaos in conservative systems.
The criterion states that large-scale chaos arises in the phase space of conservative systems when domains of resonant motion caused by separate Fourier components of a small pertubation overlap with one another. 
The criterion was first applied in celestial mechanics by \citet{Wisdom80}, who used it to predict the onset of chaos as a function of perturber-particle separation in the planar circular-restricted three-body problem.
The resonance overlap criterion has since been applied to explain the origin of chaos in a wide variety of other celestial mechanics contexts \citep[e.g.,][]{Holman:1996tu,HolmanMurray1997,Murray:1999ff,NesvornyMorbidelli1998,MudrykWu2006,Quillen2006,Mardling2008,LithwickWu2011,Quillen2011,Deck2012,Deck13,Batygin:2015ej}.
Chaos and resonance overlap can often be conveniently studied with discrete-time area-preserving dynamical maps designed to approximate sequential ``snapshots'' (i.e., Poincare sections) of the continuous-time system \citep{LLBook}.
Such maps have been applied to study $N$-body dynamics in a variety of contexts \citep[e.g.,][]{Wisdom1982,PetroskyBroucke88,DQT89,Malyshkin:1999,Pan2004} and we adapt similar methods to study the dynamics of test particles perturbed by \PX~and Neptune in this paper.

We begin, in Section \ref{SECN:NUMERICAL}, by conducting long-term $N$-body simulations consisting of a coplanar Neptune and \PX~that reproduce the anti-alignment of ESDOs observed in \BBa~and \BBb.  
In Section \ref{SECN:SECULAR}, we compare  our numerical simulations to the predictions of secular theory.
In Section \ref{SECN:CHAOTICWEB}, we examine in detail 
the resonant and chaotic dynamics of test particles orbiting under the gravitational influence of the putative \PX, as well as Neptune.
In Section \ref{SECN:DISC}, we discuss the specific implications of our results for the \PX~scenario.
%%%%%%%%%%%%%%%%%%%%%%%%%%%%%%%%%%%%%%%%%%%%%%%%%%%%%%%%%%%%%%%%%%%%%%%%%
\section{Long-term Numerical Simulations}
\label{SECN:NUMERICAL}
\subsection{Fiducial Simulation} 
\label{SECN:NUMERICAL:OVERVIEW}
As described in our introduction, one of the key outcomes of the orbital integrations undertaken by \BBa~and  \BBb, was that introducing a distant, eccentric \PX~into their simulations caused a clustering, at late times, of test particle orbits into a configuration that was anti-aligned with the \PX~orbit. 
We provide an example of this from our own simulations. 

We have conducted coplanar simulations in which Neptune is placed on circular orbit at $a=30\au$ and is accompanied by \PX~on a $a=500\au$, $e=0.6$ orbit. 
In our fiducial simulation, we take \PX's mass to be $10~\Mearth$.
In Section \ref{SEC:OTHERMASSES}, we compare our fiducial case to two additional simulations: one in which Neptune is replaced by a quadrupole approximation of its orbit-averaged contribution to the gravitational potential, and one in which  \PX's mass is reduced to $1~\Mearth$.
We initialized 3000 test-particles in orbits with $a \in [{150,550}]\au$ and 
$q\in [{33,50}]\au$  and random mean longitudes. Particles' longitudes of perihelia, $\varpi$, are chosen to be either exactly aligned ($\Delta\varpi\equiv\varpi_{P9}-\varpi=0$) or anti-aligned ($\Delta\varpi=\pi$) with \PX's periapse direction, $\varpi_9$.
In other words,  the test-particles are placed in a  population similar to those used by \BBa~and \BBb.
We integrate our systems using the IAS15 integrator \citep{RS15} in the {\sc REBOUND} code of \citet{RL12}.

\begin{figure}[htp]
    \centering
    \includegraphics[width=\columnwidth]{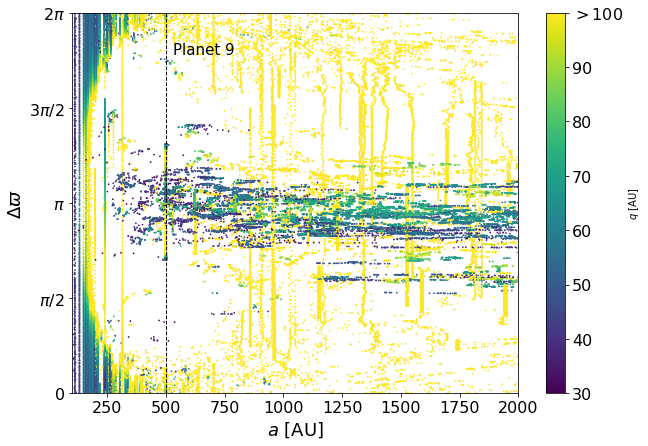}
    \caption{%
        Surviving particles from a long-term simulation of test-particle stability when orbiting in the field of both Neptune ($a\sim30\au,e=0$) and \PX~($a\sim500\au,e\sim0.6,m=10\Mearth$).
        All planets and particles are strictly coplanar.
        After $5$ Gyrs years of integration, we plot the final $2.5$ Gyrs years of data,
        illustrating that, for particles with distant orbits ($a\gsim250\au$) there is a considerable excess of points that are anti-aligned with the orbit of \PX, qualitatively similar to the results of \citet{Batygin.2016} and \citet{Brown.2016}.
        Test particle periapse distances are indicated by colorscale.
        \PX's semimajor axis is indicated with a vertical line.
    }
    \label{FIG:NUMERICAL:a_Dvarpi}
    \end{figure}

Figure \ref{FIG:NUMERICAL:a_Dvarpi} shows the results of our fiducial simulation,
where we plot the final $2.5$ Gyrs of data for particles that survive the full $5$ Gyrs of integration. Only $\sim 8\%$ of the initial 3000 particles survive the full integration and are plotted.  
Two distinct dynamical populations are evident. 
The first population, consisting of particles interior to $a\sim250\au$, evolves at approximately fixed semimajor axes with either apsidally aligned or circulating orbital orientations relative to \PX.
This interior aligned population exhibits clear structure in the  $a$--$\Delta \varpi$ space that correlates with variations in perihelion distance.  
In Section \ref{SECN:SECULAR}, we demonstrate that this structure is well-explained by secular theory. 
The aligned population consists entirely of particles that are initialized with $\Delta\varpi=0$. Some additional particles initialized with $\Delta\varpi=0$ survive the full simulation after scattering to semimajor axes $a>500\au$ and maintaining high pericenter distances. 
These particles appear as the vertical yellow lines seen at large $a$ values in Figure \ref{FIG:NUMERICAL:a_Dvarpi}.
If TNOs existed on such orbits, their large heliocentric distances would render them essentially unobservable. We devote little time to these particles in the rest of this paper because they are unimportant for explaining current observations.

The second population, by contrast, consists of particles on distant orbits ($a\gsim250\au$) that maintain anti-alignment with the orbit of \PX~($\Delta\varpi\sim\pi$) and low pericenter distances $q<100\au$ while wandering in semimajor axis. In our simulations, this population consists entirely of particles that are initialized with $\Delta\varpi=\pi$.
Similar populations were observed in simulations conducted by \BBa~and \BBb. This second population of anti-aligned particles is the fundamental means by which they explain the apparent orbital clustering of observed ESDOs.
As noted by \BBa, the dynamical evolution of the anti-aligned population, specifically their stochastic wandering in semimajor axis,  is indicative of chaotic diffusion through a web of overlapped resonances.
In Section \ref{SECN:CHAOTICWEB}, we examine the structure of this chaotic web in detail by illustrating how the mixture of resonant and chaotic trajectories are situated in the phase-space of the particles. We also address the mechanism by which the chaotically scattering particles maintain their persistent anti-alignment.
    % ------------------------------------------------------------------
    % ------------------------------------------------------------------

%%%%%%%%%%%%%%%%%%%%%%%%%%%%%%%%%%%%%%%%%%%%%%%%%%%%%%%%%%%%%%%%%%%%%%%%%
\subsection{Resonances} 

\label{SECN:NUMERICAL:RESONANCES}
\begin{figure}
        \centering
        \includegraphics[width=\columnwidth]{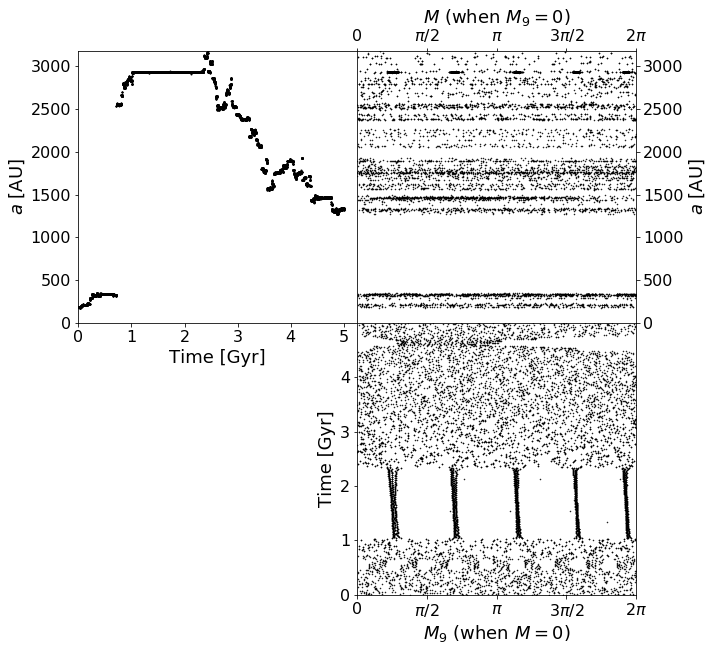}
        \caption{
        Example evolution of a test particle initially anti-aligned with \PX's orbit, from our numerical simulation.
        Each panel plots a different pairwise combination of time, the test particle's semimajor axis,
        and \PX's mean anomaly when the test particle is at pericenter.
        The test particle intermittently `sticks' in MMRs, indicated by periods of evolution where the test particle's semimajor axis remains nearly constant. During these periods, the angle $M_9$ does not densely cover the full interval $[0,2\pi)$, reflecting the phase-protection provided by libration of a resonant angle.
        }
        \label{fig:res_evolve_example}
    \end{figure}

\citet{Malhotra.2016} have noted that the periods of the four ESDOs, Sedna, 2010 GB174, 2004 VN112, and 2012 VP113, could place them in a series of $N/1$ and $N/2$ resonances with a putative \PX~at a semimajor axis $a_9\approx665\au$. \citet{Millholland17} find that a similar \PX~semimajor axis leads to commensurabilities with a larger set of ESDOs. 
However, we note that, in our simulations, essentially all of the test particles initialized in the anti-aligned configuration experience significant semimajor diffusion for the full duration of the simulation (evident from the horizontal striations in the anti-aligned regions of Figure \ref{FIG:NUMERICAL:a_Dvarpi}), and therefore are \emph{not} found to occupy any long-term stable librating resonant configurations.
Based on our numerical integrations, a significant population of long-term stable, resonant ESDOs seem unlikely to arise without dissipative forces that would aid permanent resonant capture. 
However, we often observe ``resonance sticking" where particles temporarily occupy orbits in or near resonances in both the aligned and anti-aligned populations.
Figure \ref{fig:res_evolve_example} illustrates resonance sticking for a test particle that, in the course of its chaotic evolution, is temporarily captured in various resonances. The figure plots the semimajor axis evolution of the particle, along with that of \PX's mean anomaly, $M_9$, calculated when the test-particle comes to pericenter (i.e., the test-particle mean anomaly is $M=0$).
When there is an $N$:$k$ resonance between the test-particle and \PX, this angle librates about $k$ discrete values when the particle is at pericenter.
Otherwise, outside of resonance, the angle will uniformly fill the interval $[0,2\pi)$.
The test-particle in Figure \ref{fig:res_evolve_example} is temporarily captured in the 5:9 interior MMR from 0.5 to 0.7 Gyr, the 71:5 exterior MMR from approximately 1-2 Gyr, and the exterior 5:1 MMR around 4.5 Gyr.

Temporary resonance occupation has been seen, by various other authors, in simulations of TNOs in the presence of \PX.
As noted in the introduction, \citet{BatyginMorbidelli17} observe resonance-hopping  behavior when  particles are given small orbital inclinations relative to \PX.
However, the test-particles in the simulations of \citet{BatyginMorbidelli17}, which do not fully account for Neptune's gravitational influence, appear to spend a larger fraction of their orbital evolution in resonances when compared with our fiducial simulation.
\citet{Becker2017} observe resonance-hopping behavior in simulations of observed TNOs influenced by Neptune, \PX~(which, in their simulations, is inclined $30^{\circ}$ to the ecliptic), and the quadrupole potential of the other outer solar system planets.
In simulations from \citet{Becker2017}, TNOs experience significant migration (changes in semimajor axis of $\Delta a>100\au$) roughly as often as they stay close their initial semimajor axis while residing in a single resonance or exhibiting small jumps between neighboring resonances.
\citet{Millholland17} find persistent diffusion of semimajor axes for most/all particles and hence only temporary residence near any given MMR in their numerical simulations that include Neptune's full gravitational potential and the quadrupole potential of the other giant planets (S. Millholland, private communication).

    % ----------------------
    % ----------------------

\subsection{Other Simulations}
\label{SEC:OTHERMASSES}
\begin{figure}[htp]
    \centering
    \includegraphics[width=\columnwidth]{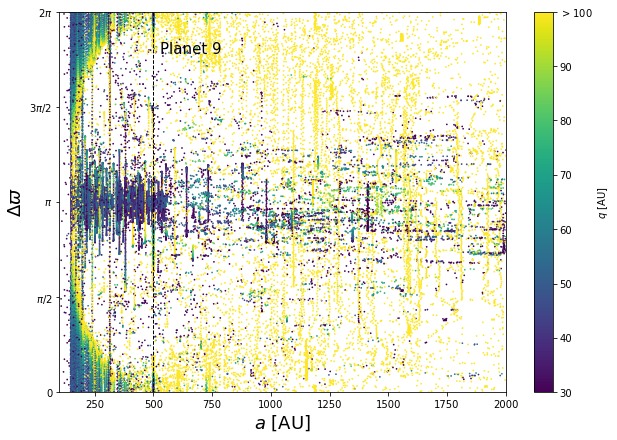}
    \caption{Same as Figure \ref{FIG:NUMERICAL:a_Dvarpi} but for a simulation in which Neptune is replaced by a quadrupole potential.}
    \label{FIG:NUMERICAL:J2_a_Dvarpi}
    \end{figure}

\begin{figure}[htp]
    \centering
    \includegraphics[width=\columnwidth]{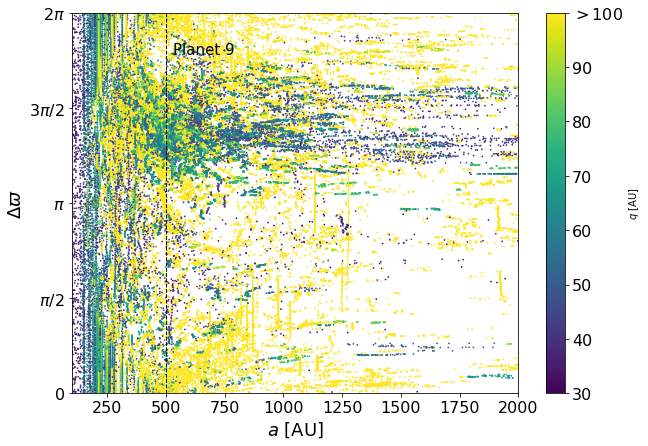}
    \caption{Same as Figure \ref{FIG:NUMERICAL:a_Dvarpi} but for a simulation with a $1\Mearth$ \PX.}
    \label{FIG:NUMERICAL:1MEARTH_a_Dvarpi}
    \end{figure}

In addition to our fiducial simulation, we have run one in which Neptune is replaced by its orbit-averaged quadrupole contribution to the gravitational potential and one with a $1\Mearth$ \PX.
Hereafter, we will refer to these as the ``quadrupole" and ``low-mass" simulations, respectively.
The results of these simulations are shown in Figure \ref{FIG:NUMERICAL:J2_a_Dvarpi} and \ref{FIG:NUMERICAL:1MEARTH_a_Dvarpi}.
Both show significant differences from the fiducial simulation presented in Section \ref{SECN:NUMERICAL:OVERVIEW}.

Approximately $\sim 1/3$ of the particles survive up to 5 Gyr in the quadrupole simulation (Figure \ref{FIG:NUMERICAL:J2_a_Dvarpi}), significantly more than the $\sim 8\%$ survival fraction of the fiducial simulation.
Significantly more anti-aligned particles remain at semimajor axes interior to \PX~in the quadrupole simulation than in the fiducial simulation. 
These particles maintain constant semimajor axes for most or all of the simulation, causing the vertical striations seen in Figure \ref{FIG:NUMERICAL:J2_a_Dvarpi} near $\Delta\varpi\approx \pi$ and $a\lesssim500\au$. In contrast, anti-aligned population members undergo significant chaotic semimajor axis evolution in the fiducial simulations.

\citet{BatyginMorbidelli17} investigate the dynamics of test-particles under the gravitational influence of \PX~and a quadrupole potential representing the solar system's giant planets. 
Thus, their simulations are qualitatively similar to our quadrupole simulation, and much of their analysis readily explains the results of our quadrupole simulation. 
In particular, they show that:
\begin{enumerate}
    \item Anti-aligned particles survive the duration of the simulation in MMRs with \PX~that protect them from close encounters.
    \item The secular evolution of these resonant particles maintains their anti-alignment with \PX.  \citet{BEUST16} also demonstrated that resonant particles are expected to maintain anti-alignment over the course of their secular evolution in the presence of \PX~and the quadrupole potential of the outer solar system.
\end{enumerate}
The quadrupole simulations adequately capture the features of the circulating/aligned population of test-particles seen in the fiducial simulation. 
This is because, as we show below in Section \ref{SECN:SECULAR}, their dynamics are well-described by a secular approximation. 
However, the dynamics of the anti-aligned population are significantly more chaotic in the fiducial simulation. \citet{BatyginMorbidelli17}'s explanation of the dynamical mechanism maintaining anti-alignment---secular evolution in resonance---appears to be incomplete: the inclusion of Neptune's full gravitational potential disrupts nearly all stable resonant librations in the fiducial simulation. 
Nonetheless, the fiducial simulation still exhibits a population of particles that maintain persistent anti-alignment with \PX.
In Section \ref{SECN:CHAOTICWEB:NEPTUNE}, we show that Neptune alone can induce significant chaotic behavior in particles with semimajor axes in our initial range $a\in [150,550]~\au$, even for pericenter distances as large as $q\sim 50~\au$.
Therefore, it is unsurprising that the fiducial simulation shows significantly different evolution from the quadrupole simulation.
% ------------------------------------------------------------------
% ------------------------------------------------------------------

Figure \ref{FIG:NUMERICAL:1MEARTH_a_Dvarpi} shows the results of our low-mass simulation.
As in the fiducial simulation of Figure \ref{FIG:NUMERICAL:a_Dvarpi}, the test-particles can roughly be separated into two populations: a `circulating/aligned' population at small semimajor axes and an `anti-aligned' population that experiences chaotic semimajor axis evolution.
The most striking difference is that the `anti-aligned' population is no longer directly anti-aligned with \PX's orbit, but rather appears to be centered around $\Delta\varpi\sim 3\pi /2$. In Section \ref{SECN:CHAOTICWEB:APSIDAL}, we argue  that this concentration arises because the distribution of perihelion distances and longitudes does not relax to a steady state over the course of our 5 Gyr simulation.

We conducted some preliminary numerical simulations with \PX's mass increased to 100$\Mearth$.
These simulations produce orbital clustering similar to what is observed in the fiducial simulation, albeit with substantially fewer surviving test particles. 
We do not consider this high-mass simulation further because observational constraints probably rule out such a large \PX~mass \citep{Luhman.2014}.

We have also run a simulation identical to the fiducial simulation presented in Section \ref{SECN:NUMERICAL:OVERVIEW} but with each test particle given a random initial $\varpi$, rather than starting from perfectly aligned and anti-aligned configurations.  We find that only 58 particles out of an initial population of 1500 (i.e., $\sim4\%$) survive the full simulation when started from random initial apsidal alignments.
{Randomizing initial apsidal alignments reduces the particles' survival rate, because it} places many more particles on secular trajectories that evolve through the crossing/non-crossing boundary (see Section \ref{SECN:SECULAR} below).
As we will see in Section \ref{SECN:CHAOTICWEB}, particles experience strong resonance overlap as they approach this boundary, and they tend to be ejected. 
We plan to explore the consequences of randomizing initial particle alignment more thoroughly in a forthcoming paper \citep{Li_inprep}. 

% ------------------------------------------------------------------
% ------------------------------------------------------------------

\subsection{Summary} 
\label{SEC:SIMSUMMARY}
To summarize, the key features of our long-term simulations (which we seek to explain below in Sections \ref{SECN:SECULAR} and \ref{SECN:CHAOTICWEB}) are:
\begin{enumerate}
\item{Surviving particles inside  $a\lesssim 250\au$ have apsidal lines that circulate or maintain alignment with \PX's. These particles in this aligned population do not experience large excursions in semimajor axis. }
\item{ The majority of surviving particles outside $\sim 250\au$ in the fiducial simulation are anti-aligned with \PX's orbit. These particles maintain their anti-alignment over the course of the simulation while diffusing in semimajor axis.}
\item{Nearly all surviving anti-aligned particles were {\it initialized} with an anti-aligned configuration. Likewise, the aligned population is composed of particles initially in an aligned configuration.}
\item{Particles often  temporarily ``stick" to MMRs, but no particles survive the full simulation librating stably in a single resonance. This is in contrast to the "quadrupole" simulation without an active Neptune, where many particles are observed to stably librate in various MMRs with \PX~for the duration of the simulation.}
\item{Substantially more test particles are retained when \PX's mass is decreased to $1\Mearth$. The population of test particles can again be divided into two separate populations: one with circulating or aligned longitudes of perihelia that remain at roughly constant semimajor axis, and a second population that undergoes chaotic semimajor axis diffusion. The orbital alignments of the  chaotically scattering population cluster around $\Delta\varpi\sim3\pi/2$ rather than $\Delta\varpi\sim\pi$ as observed when \PX~has a mass of $10\Mearth$.}
\end{enumerate}
% In Section \ref{SECN:CHAOTICWEB} we detail the \emph{mechanisms} which drive this asymmetric population of anti-aligned distant orbits. 
%%%%%%%%%%%%%%%%%%%%%%%%%%%%%%%%%%%%%%%%%%%%%%%%%%%%%%%%%%%%%%%%%%%%%%%%%
\section{Secular dynamics}
\label{SECN:SECULAR}
    % ------------------------------------------------------------------
    \begin{figure*}[htp]
    \centering
    \includegraphics[width=0.4\textwidth]{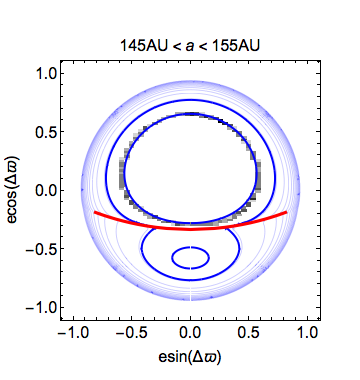}
    \includegraphics[width=0.4\textwidth]{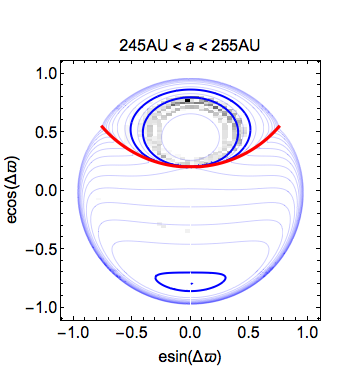}
    \includegraphics[width=0.4\textwidth]{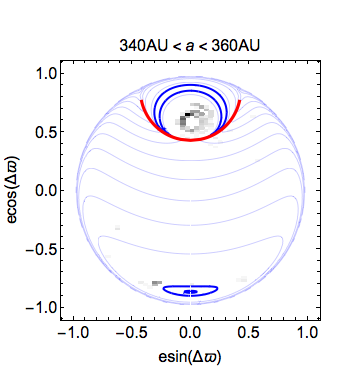}
    \includegraphics[width=0.4\textwidth]{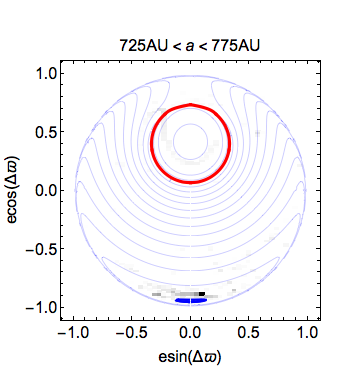}
    \caption{%
    Test particle density maps (grayscale) in the $(e\sin{\Delta\varpi},e\cos{\Delta\varpi})$ plane from long-term simulations when orbiting in the field of both Neptune and \PX, overlaid with constant energy contours of the secular averaged problem (thin blue lines).
    Particles aligned with \PX~are located toward the \emph{top} of each panel (anti-aligned at the bottom).
    Each slice is at a fixed \emph{range} of $a$ indicated above each panel, and the secular contours are computed for the semimajor axis value at the midpoint of this range. 
    The red line in each panel separates the orbits that are \PX-crossing (below the line) from those that are not (above).
    Dark blue contours indicate trajectories that reach a perihelion distance $q=35$~and $50$\au, corresponding to the minimum and maximum initial perihelion distances in the numerical simulations. 
    }
    \label{FIG:NUMERICAL:LONG_DENSITY}
    \end{figure*}
    % ------------------------------------------------------------------

Figure \ref{FIG:NUMERICAL:LONG_DENSITY} shows the distribution of surviving test particles' orbital eccentricities and alignments in coordinates $\{e\cos\Delta\varpi,e\sin\Delta\varpi\}$ over the course of the simulation in four different semimajor axis ranges.
The distribution % of the surviving particles' eccentricities and alignment 
closely follows the shape of constant secular energy levels, which are over-plotted. (Appendix \ref{SECN:APP:Equation} provides the details of how secular energy contours are computed in \figref{FIG:NUMERICAL:LONG_DENSITY}.)
The morphology of the secular energy contours in Figure \ref{FIG:NUMERICAL:LONG_DENSITY} provides insight into the transition from alignment to anti-alignment among the surviving particles at a critical semimajor axis of $a_\text{crit}\sim 250$\au.

\subsection{Aligned Particles}\label{S:S:A}
Inside $a_\text{crit}$, the secular evolution of most aligned particles from their initial configuration with perihelia in the range $q\in[35\au ,50\au]$ follow trajectories that librate about $\Delta\varpi=0$ and avoid evolving onto crossing orbits.
% These trajectories are highlighted in \figref{FIG:NUMERICAL:LONG_DENSITY} by 
The disappearance of the aligned population beyond $a_\text{crit}$ occurs where the secular trajectories on which aligned particles are initialized all evolve onto crossing orbits.\footnote{Some aligned particles that have experienced chaotic scattering are present beyond $a_\text{crit}$  in the  bottom left $340~\au<a<360~\au$ panel of Figure \ref{FIG:NUMERICAL:LONG_DENSITY}. 
These particles are no longer on secular trajectories in the range corresponding to the simulation's initial conditions. Their new  secular trajectories avoid nearly crossing orbits with \PX~that would destabilize them.}
This can be seen in Figure \ref{FIG:NUMERICAL:LONG_DENSITY} where trajectories starting in the initial range terminate at the red line separating crossing from non-crossing orbits, rather than closing on themselves above this line.
Orbits on the red crossing/non-crossing line meet \PX's orbit tangentially.  Orbits near this line are prone to close encounters with \PX. 
The strength of these close encounters is greatly enhanced by the fact that they occur near the test particle's apoapse, when the particle is moving most slowly. These encounters tend to eject the test particles.
In Section \ref{SECN:CHAOTICWEB}, we more thoroughly examine the onset of chaos induced by resonance overlap as the orbit-crossing line is approached.
%
%%%%%%%%%%%%%%%%%%%%%%%%%%%%%%%%%%%%%%%%%%%%%%%%%%%%%%%%%%%%%%%%%%%%%%%%%

\subsection{Anti-aligned Particles}\label{S:S:AA}
Essentially all particles in the anti-aligned population experience chaotic semimajor axis evolution; hence, a purely secular treatment is \emph{not} strictly applicable to their dynamics.
Nonetheless, the morphology of anti-aligned libration islands appears to match the distribution of anti-aligned orbits.
The anti-aligned particles are concentrated near the secular trajectories that correspond to their initial periapse distances. 
Few anti-aligned particles exist in the semimajor axis range ($145~\text{\au}<a<155~\text{\au}$) shown in first panel of Figure \ref{FIG:NUMERICAL:LONG_DENSITY}.  
In this range, the initial secular trajectories evolve toward the boundary from crossing to non-crossing.
Resonance overlap is strongest near this transition, as described previously and demonstrated below in Section \ref{SECN:CHAOTICWEB}.
We will return to the question of the anti-aligned population's apsidal confinement mechanism in Section \ref{SECN:CHAOTICWEB:APSIDAL}.
%, so that particles approaching this boundary are ejected. This can be understood intuitively from the fact near this boundary particles are subject to prolonged close encounters with \PX~near their aphelion.

%%%%%%%%%%%%%%%%%%%%%%%%%%%%%%%%%%%%%%%%%%%%%%%%%%%%%%%%%%%%%%%%%%%%%%%%%
\section{MMRs and the Chaotic Web}
\label{SECN:CHAOTICWEB}
In this section, we chart the chaotic web of overlapped MMRs with Neptune and \PX, and relate its structure to the long-term evolution of test particles in our numerical simulations. 
In Section \ref{SECN:CHAOTICWEB:ALIGNED}, we consider the role of chaos in determining where the aligned population can remain on stable orbits.
Section \ref{SECN:CHAOTICWEB:ANTI} provides an overview of the mixture of chaotic and regular phase-space inhabited by anti-aligned particles with short-term numerical simulations.
In Section \ref{SECN:CHAOTICWEB:RESDYNAMICS}, we consider the dynamics of a single resonance for an orbit-crossing anti-aligned particle. We describe how these resonances and their overlap shape the global phase space of anti-aligned particles in Section \ref{SECN:CHAOTICWEB:OVERLAP}.
In Section \ref{SECN:CHAOTICWEB:NEPTUNE}, we consider Neptune's contribution to the chaotic evolution of particles.
Finally, in Section \ref{SECN:CHAOTICWEB:APSIDAL}, we revisit the apsidal evolution of the chaotically diffusing anti-aligned population.
%
% In any of the following simulations which include Neptune, it is initialized on a circular orbit at a semimajor axis $a=30$ \au.
% In any of the following simulations which include \PX, it is initialized with orbital parameters $a=500$\au and $e=0.6$.
% %
% We explore a range of masses for \PX.

    % ------------------------------------------------------------------
    \begin{figure}[htp]
    \centering
    \includegraphics[width=\columnwidth]{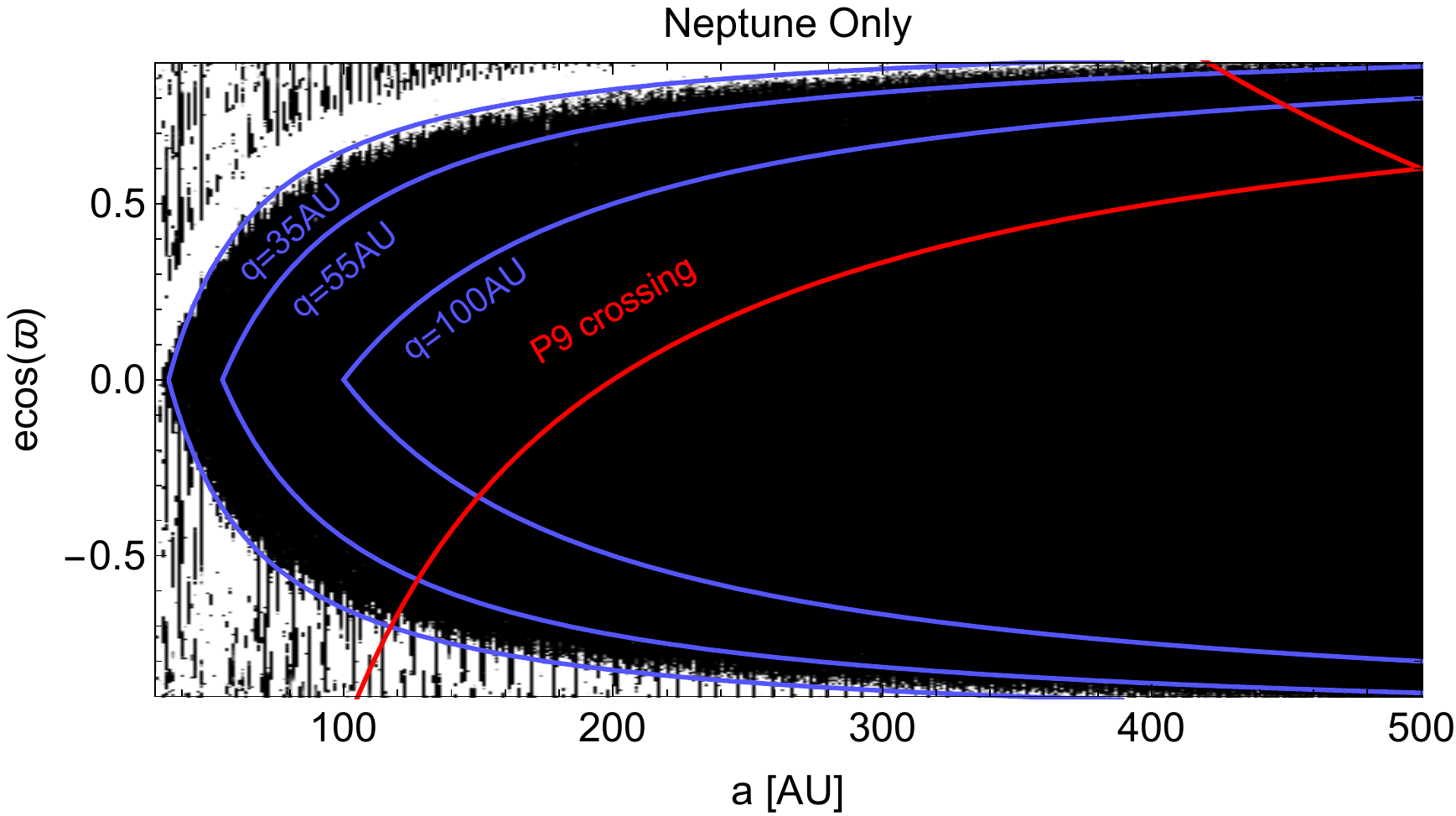}
    \includegraphics[width=\columnwidth]{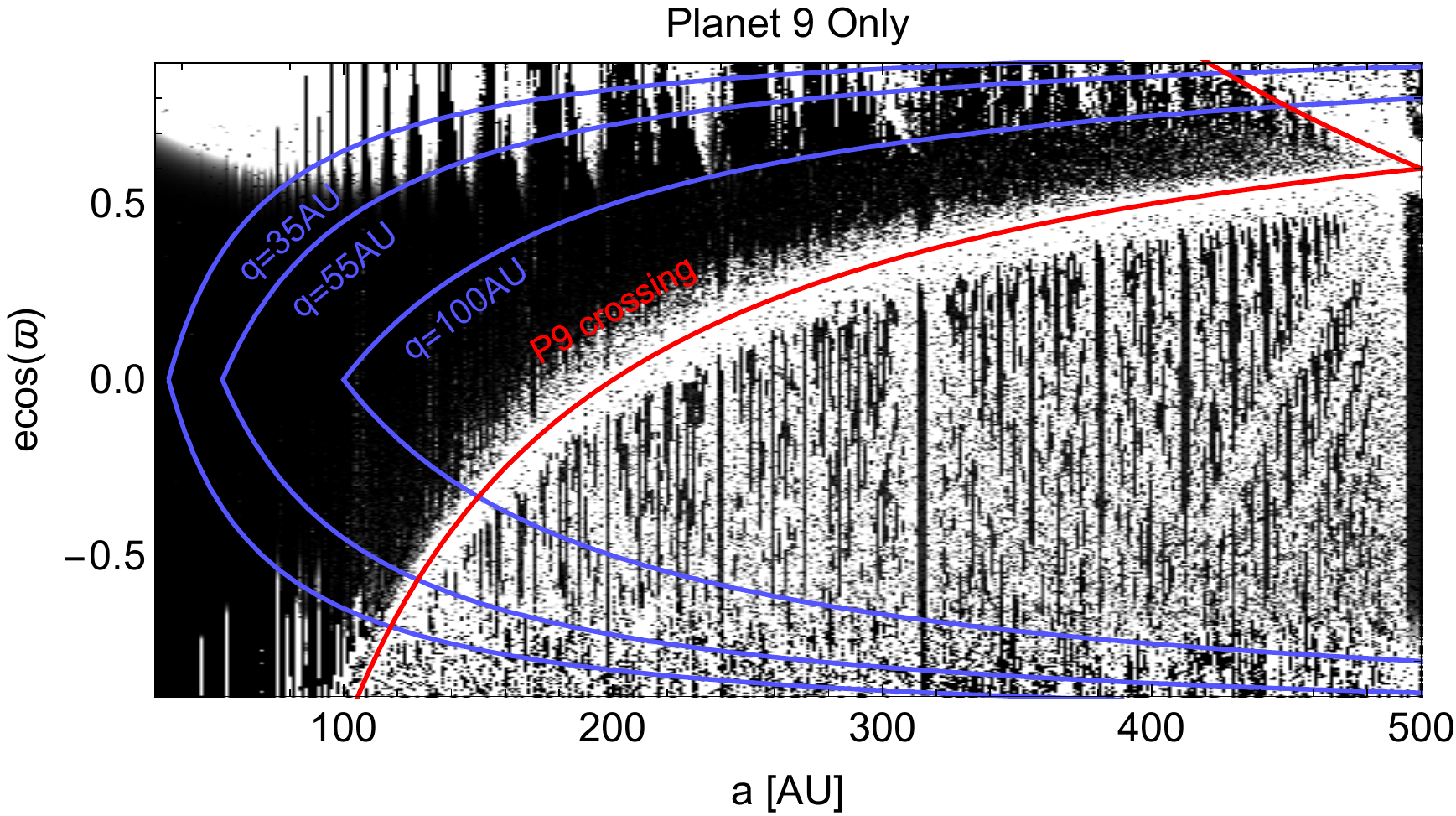}
    \includegraphics[width=\columnwidth]{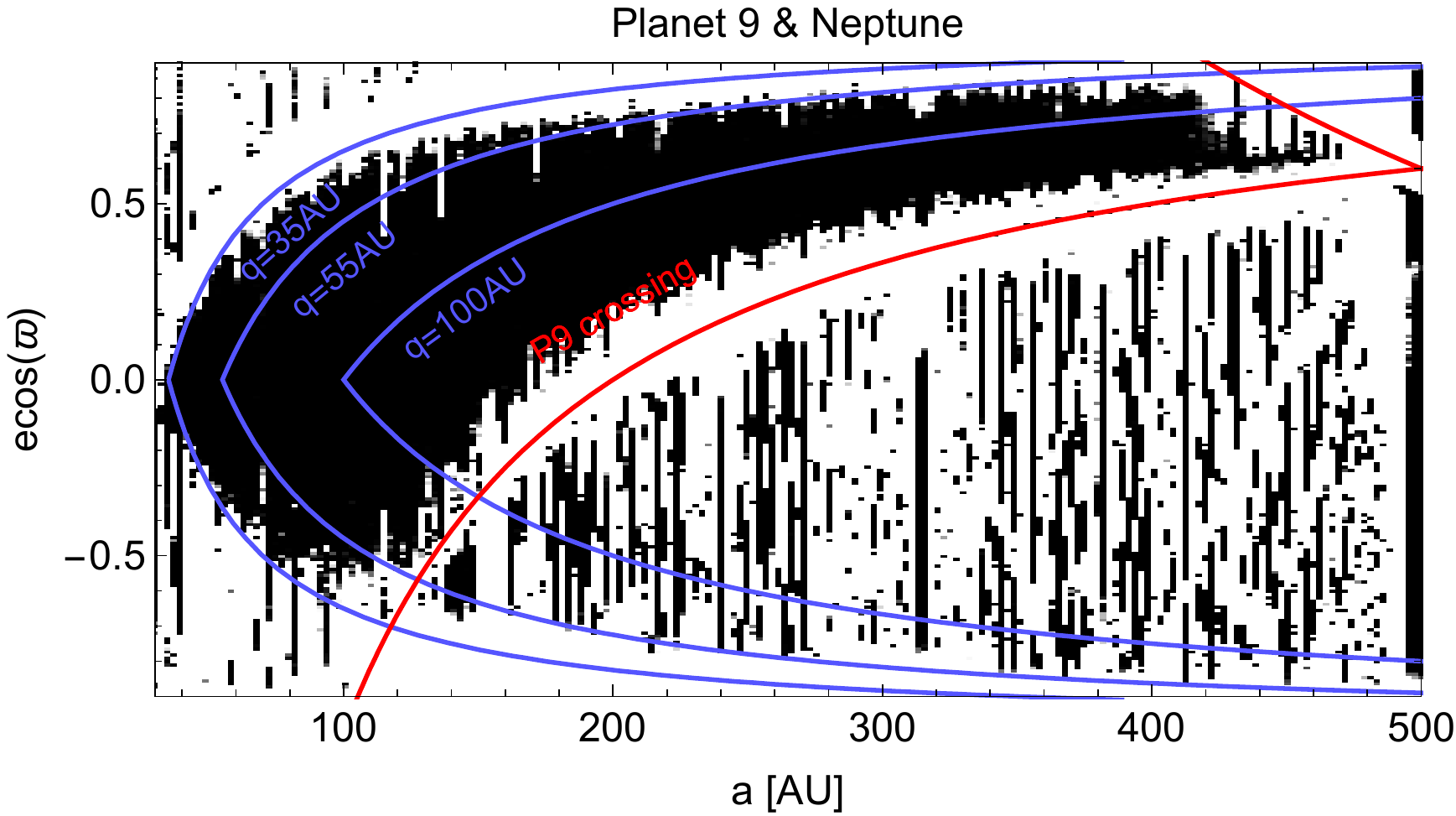}
    \caption{%
        Test-particle stability in the $(a,e\cos{(\varpi=\pm\pi)})$ plane when orbiting in the field of: 
        {(top)} Neptune only ($a_N=30$\au, $e_N=0$);
        {(middle)} \PX~only ($m_9=10\Mearth$, $a_9=500$\au, $e_9=0.6$);
        {(bottom)} Both Neptune and \PX.
        We measure the orbital alignment of all test particles, $\varpi$, with respect to the orbit of \PX.
        Particles in the top half of the plots are initialized such that they are \emph{aligned} with the orbit of \PX~($\varpi=0$), while those in the plots are \emph{anti-aligned} with the orbit of \PX~($\varpi=\pi$).
        Lines corresponding to constant pericenter distances $q=35,~55,$ and $100$\au~are shown in blue.
        The red line indicates the boundary for \PX-crossing orbits.
    }
    \label{FIG:COP:acosw}
    \end{figure}
    % ------------------------------------------------------------------

To chart the chaotic web of overlapped MMRs with Neptune and \PX, we make use of ``stability maps" constructed by computing MEGNO chaos indicators \citep{Cincotta2003} from short-term integrations  on grids in parameter space.\footnote{{While chaos does not necessarily imply orbital instability, we find that essentially all trajectories identified as chaotic by their MEGNO values experience significant changes to their orbital elements. 
Therefore, it is appropriate to refer to these maps as ``stability maps."}} 
We do this using the MEGNO functionality within the {\sc REBOUND} code \citep{RL12}.
Figure \ref{FIG:COP:acosw} shows a series of such stability maps in the plane of the long-term simulations' initial conditions.
Grid points are colored according to their MEGNO value on a grayscale that stretches from $\text{MEGNO}=2$ (black) to $\text{MEGNO}=6$ (white) based on integrations lasting lasting 300 test-particle orbits. 
The narrow grayscale range provides a sharp distinction between trajectories that behave regularly on short timescales (black) from those that exhibit chaos (white).
From the stability maps in Figure \ref{FIG:COP:acosw} we see that large-scale chaotic regions (plotted in white) occur where test particles' orbits are (nearly) crossing either Neptune or \PX's orbit. 
The chaotic regions seen in the bottom panel, where both Neptune and \PX~are present, can mostly be attributed to one of the two separate chaotic regions associated with Neptune or \PX, shown in the top two panels. 
\subsection{Aligned particles}
\label{SECN:CHAOTICWEB:ALIGNED}
Surviving particles in the aligned population maintain relatively constant semimajor axes.  
Consequently, secular averaging accurately approximates their long-term dynamics, as we saw in Section \ref{SECN:NUMERICAL}.
However, a large fraction ({$88\%$}) of initially aligned particles are ejected from the simulation, and many of the surviving particles experience limited chaotic variations in semimajor axis. Therefore, chaos caused by resonance overlap plays some role in shaping the distribution of the aligned population.

In Section \ref{SECN:SECULAR} we saw that initially aligned particles do not survive on secular trajectories that lead to orbit-crossing with \PX. We argued that this is because these particles suffer strong close encounters with \PX~that lead to their ejection.
In fact, initially aligned particles that evolve secularly onto \emph{nearly} crossing orbits experience significant chaotic variations in their semimajor axes and are often ejected.
This is in agreement with Figure \ref{FIG:COP:acosw} where it can be seen that significant chaos occurs slightly before eccentricities reach orbit-crossing values. 
%

% ------------------------------------------------------------------
\begin{figure*}
    \centering
    \includegraphics[width=0.4\textwidth]{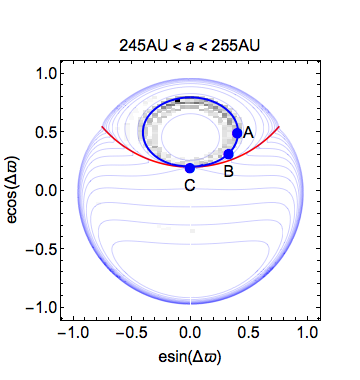}
    \raisebox{+.35\height}{\includegraphics[width=0.4\textwidth]{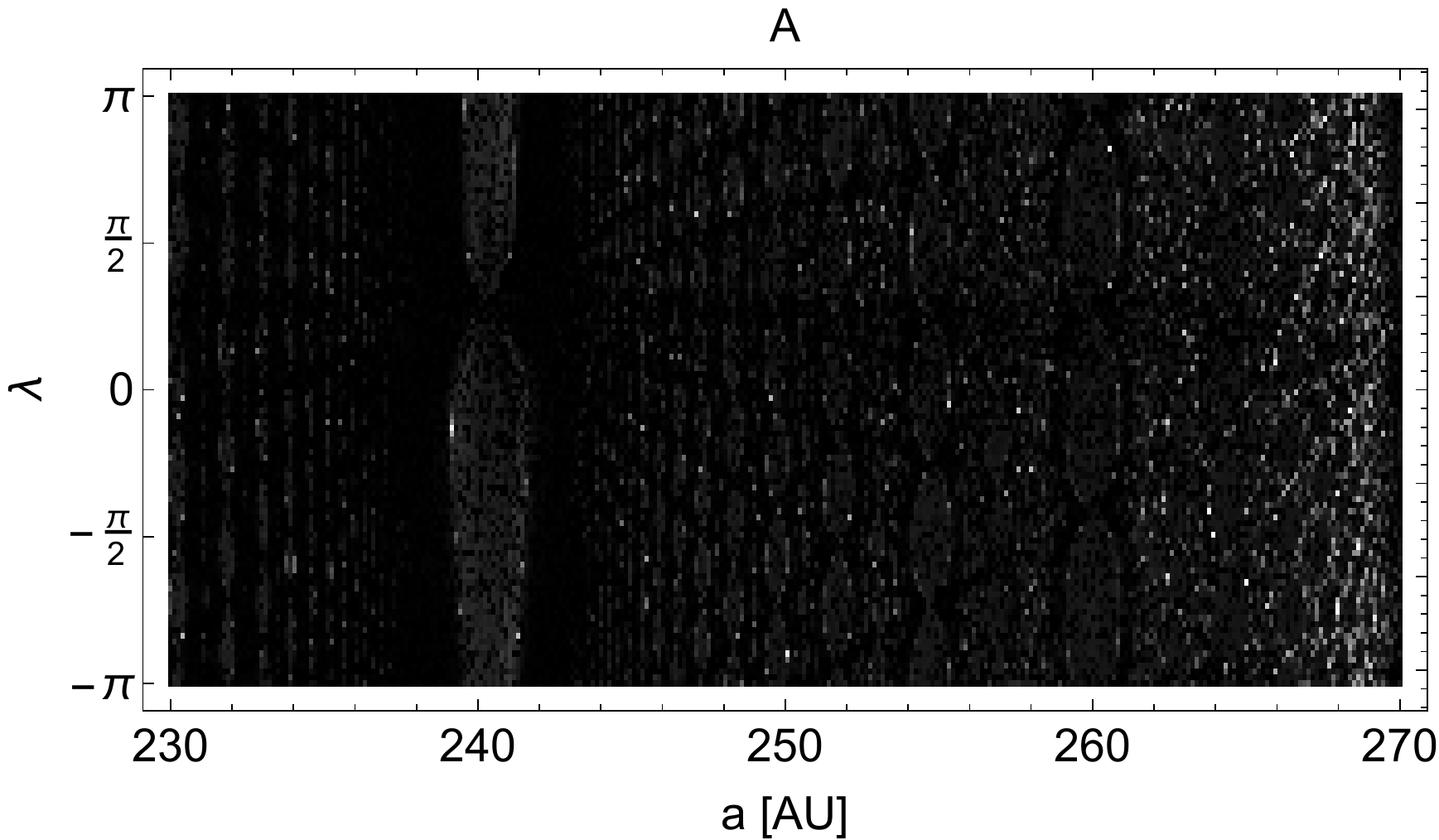}}
    \includegraphics[width=0.4\textwidth]{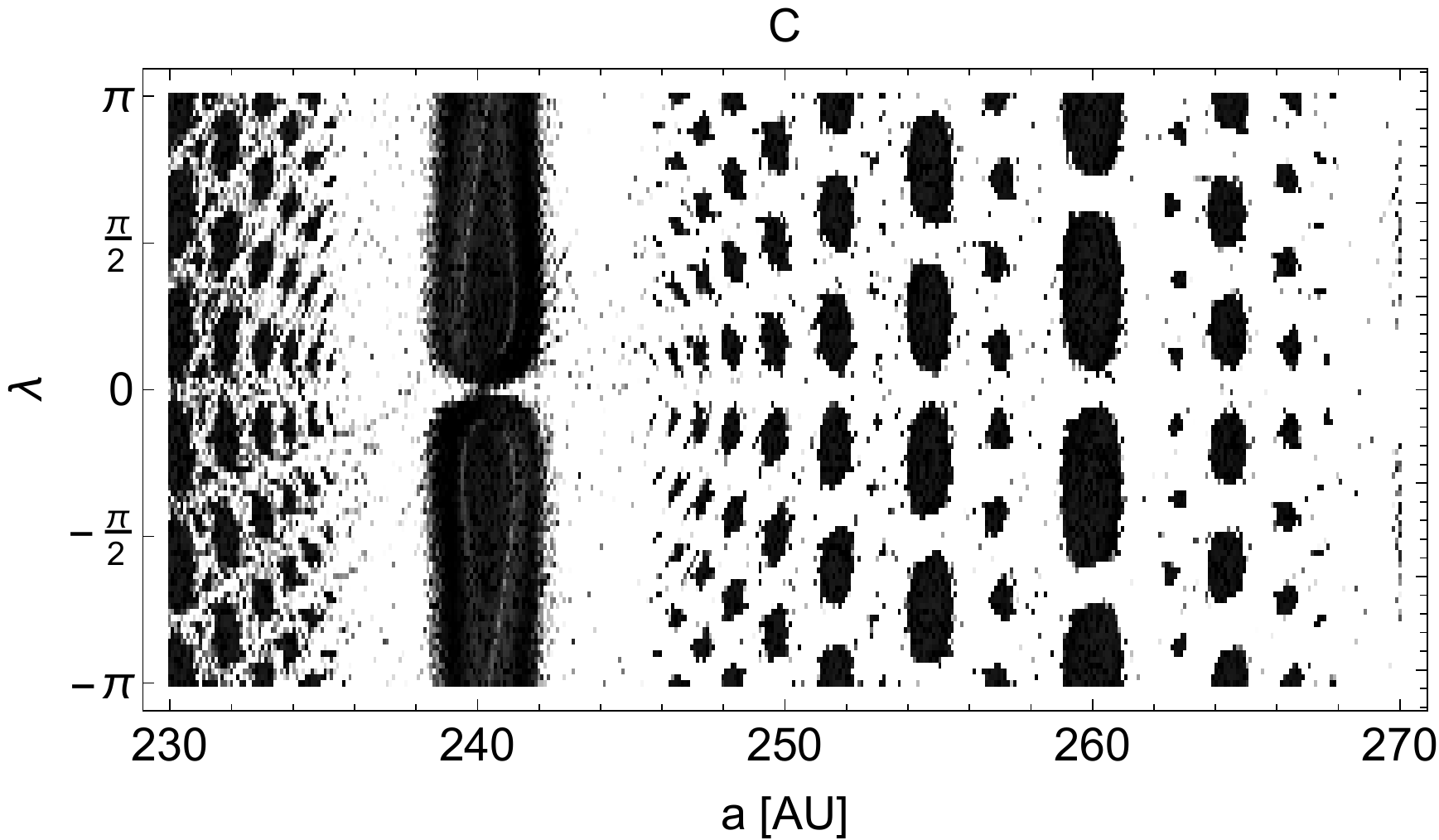}
    \includegraphics[width=0.4\textwidth]{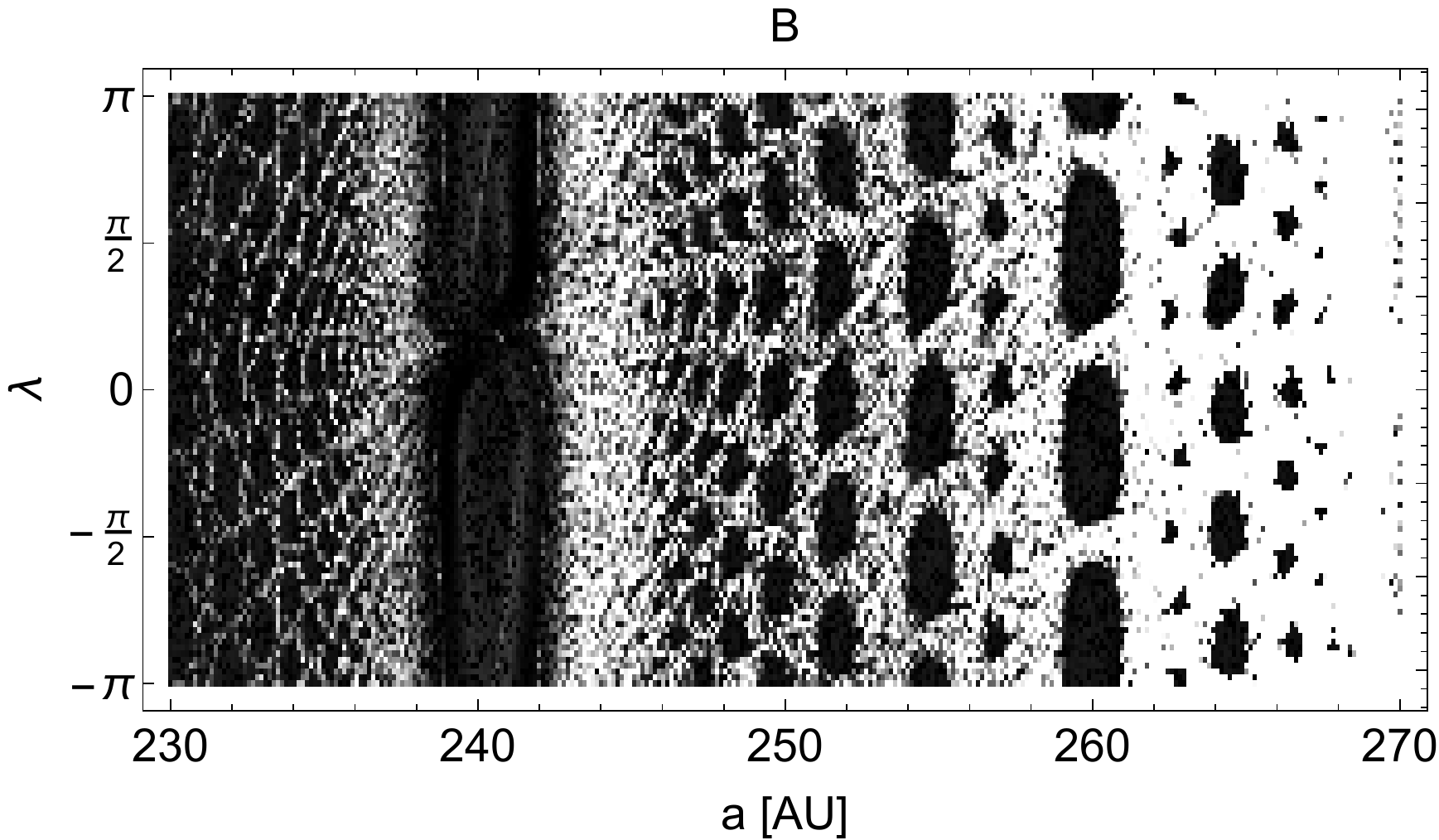}
    \caption{Short-term stability maps at different points along a  secular evolution  trajectory for a particle initially aligned with \PX. 
    The top-left panel shows a density map in $e\cos(\Delta\varpi),e\cos(\Delta\varpi)$ with secular energy contours in blue reproduced from the top-right-hand plot in Figure \ref{FIG:NUMERICAL:LONG_DENSITY}. Other panels shows stability maps in semimajor axis, $a$, and mean longitudes, $\lambda$, with particles' $e$ and $\Delta\varpi$ values initialized to the values at points "A","B", and "C"  labeled along the bold secular energy-level contour in the left-middle panel. 
    As the secular evolution proceeds sequentially from point "A" to point "C," particles evolve onto orbits that approach \PX-crossing and MMRs with \PX~become progressively more overlapped.}
    \label{FIG:CHAOTICWEB:ALIGNEDRING}
\end{figure*}
% ------------------------------------------------------------------

In Figure \ref{FIG:CHAOTICWEB:ALIGNEDRING}, we examine in detail an example of how secular evolution onto nearly crossing orbits brings particles into a chaotic region of phase space.
Figure \ref{FIG:CHAOTICWEB:ALIGNEDRING}  shows a series of stability maps in test-particle semimajor axis $a$ and mean longitude $\lambda$. 
The simulations are initialized with \PX~at its pericenter.
Also shown in \figref{FIG:CHAOTICWEB:ALIGNEDRING}, in the top-left panel, is a contour/density plot in the  $(e\sin{\varpi},e\cos{\varpi})$ plane reproduced from Figure \ref{FIG:NUMERICAL:LONG_DENSITY}, with a series of three points labeled "A" through "C" along a secular trajectory chosen from the range of our fiducial simulation's initial conditions.
Each stability map is computed by initializing the test particles' eccentricities and longitudes of perihelia to the values at one of these points along the secular trajectory. 
From  \figref{FIG:CHAOTICWEB:ALIGNEDRING},  we see how particles' orbits slowly evolve into progressively more chaotic regions of phase space.
Initially, orbits start near the top-most point on the secular trajectory shown in \figref{FIG:CHAOTICWEB:ALIGNEDRING}. 
Here, orbits behave in an entirely regular fashion on short timescales.
As secular evolution evolves orbits clockwise around the secular trajectory, they reach point "A."
The stability map corresponding to point "A" indicates mostly regular trajectories (black), with faint hints of chaos beginning to appear (white). 
Subsequent panels "B" and "C" show a progressively more chaotic phase space punctuated by regular islands associated with libration in various MMRs.
For example, there is a prominent stable black region associated with libration in the 3:1 MMR at $a\approx240\au$.
Upon reaching eccentricities and longitudes of perihelia near point "C," many of the simulation particles experience significant resonance overlap resulting in vigorous chaos and ejection. For particles that manage to survive, continued secular evolution brings their orbits back to regions of phase space where resonances are no longer overlapped. (The secular trajectory is symmetric about $e\sin\Delta\varpi=0$,  and stability maps computed with $e\sin\Delta\varpi<0$ mirror those with $e\sin\Delta\varpi>0$ but with $\lambda\rightarrow -\lambda$, so that after reaching "C," particles effectively go through the same sequence in reverse order.)

The critical semimajor axis, $a_\text{crit}$, beyond which initially \PX-aligned orbits do not survive, is the semimajor axis at which the secular trajectories bring all particles sufficiently deep into the chaotic zone near orbit-crossing that they are all ejected.
The mass of \PX~influences where this boundary occurs via two effects: first, a larger \PX~mass changes the contours of the secular Hamiltonian, and therefore the initial pericenter distances of orbits that evolve to become nearly \PX-crossing.
Second, a larger \PX~mass causes the onset of large-scale chaos to occur further from the orbit-crossing boundary.

\subsection{Anti-aligned particles}
\label{SECN:CHAOTICWEB:ANTI}
    
    \begin{figure}
        \centering
        \includegraphics[width=\columnwidth]{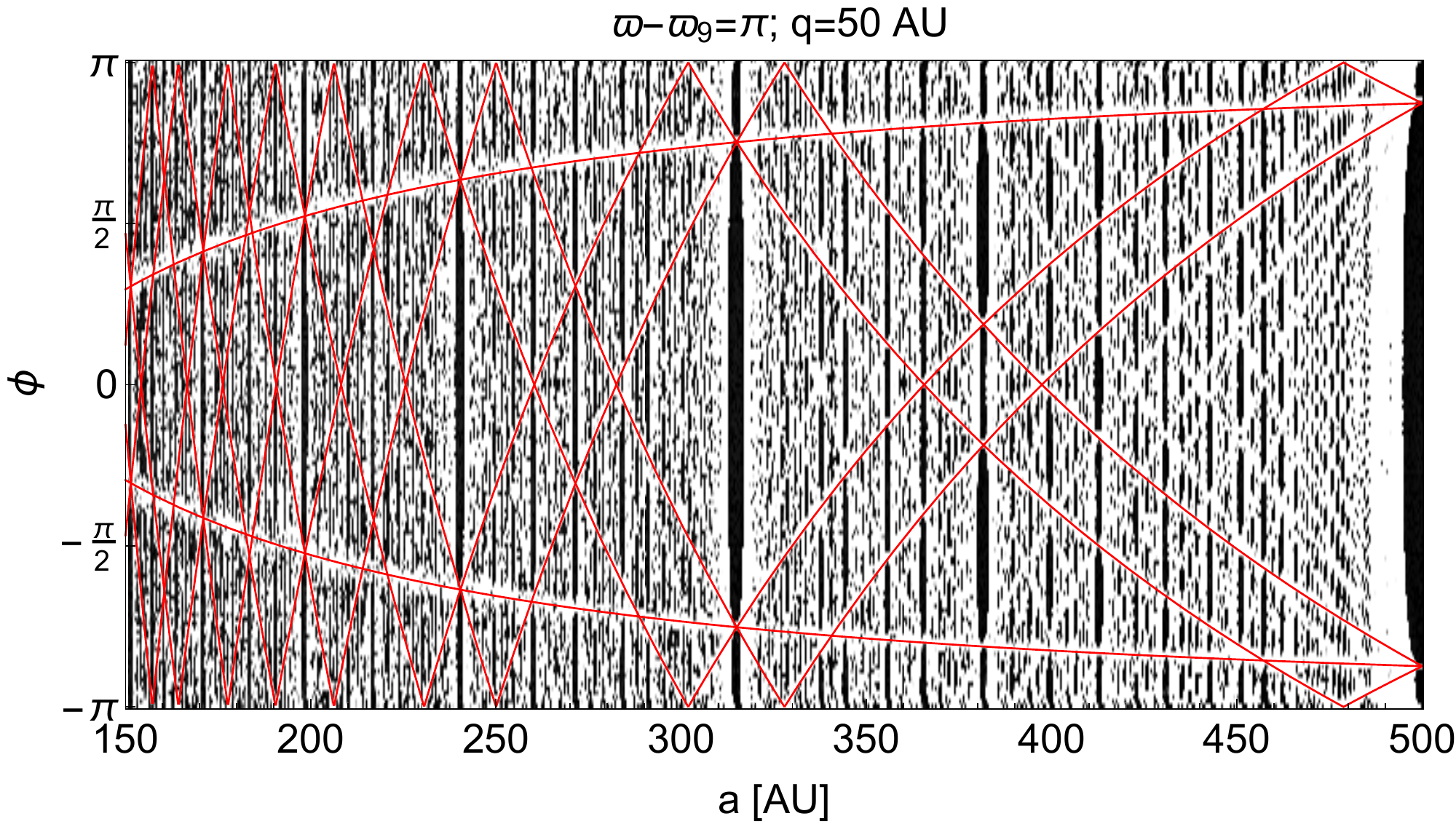}
        \includegraphics[width=\columnwidth]{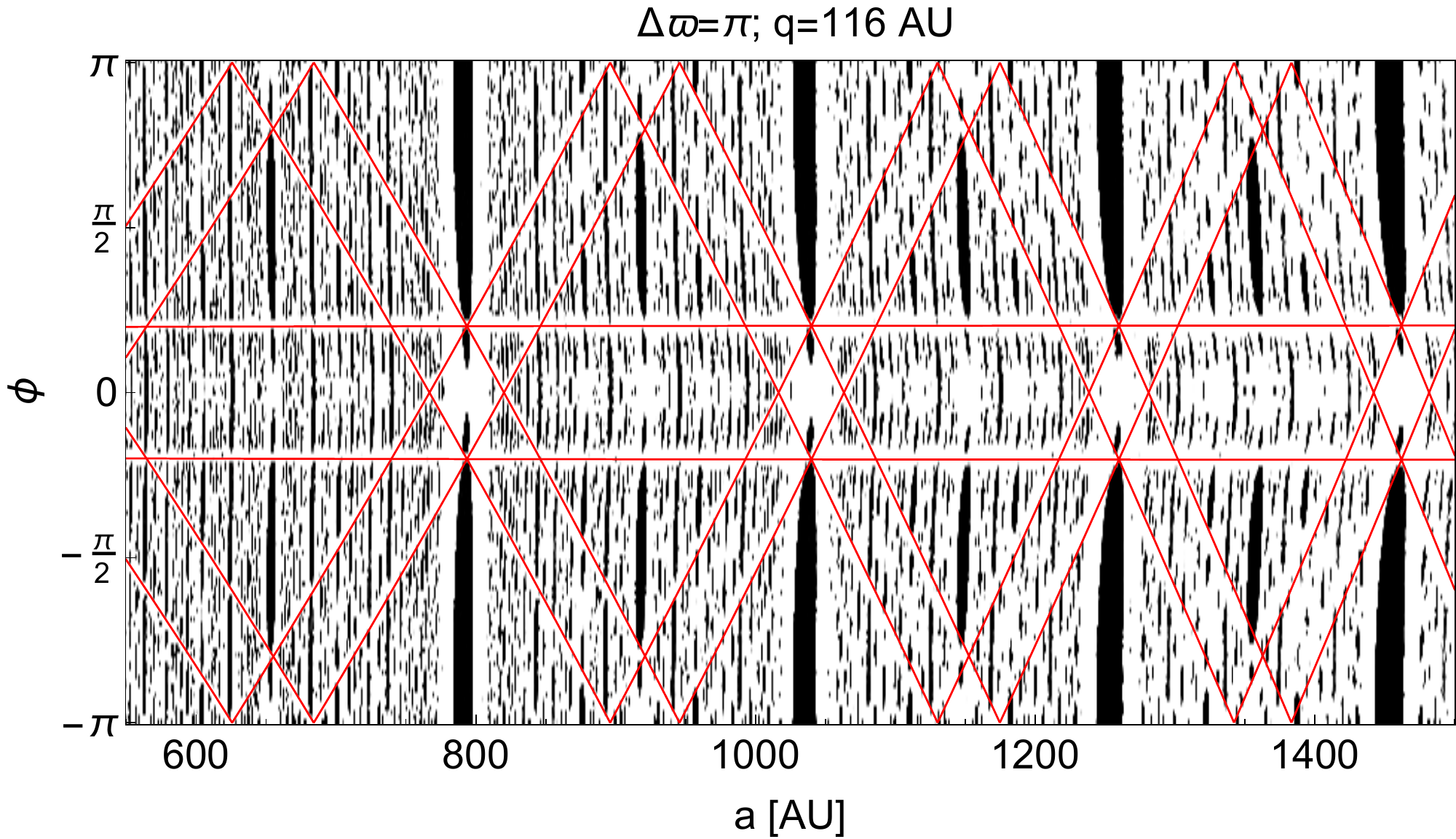}
        \caption{Stability maps illustrating resonant structure for anti-aligned particles on crossing orbits with \PX. 
        Red lines trace initial conditions that lead to a collision with \PX~within $\pm$1 orbit.
        In the top (resp., bottom) panel, \PX~(resp., the test particle) is initialized at pericenter and $\phi$ is the initial mean anomaly of the test particle (resp., \PX). In both panels, particles' orbits are initialized perfectly anti-aligned with \PX's orbit (i.e., $\Delta\varpi=\pi$).
        }
        \label{FIG:COPLANAR:WEBOUTER}
    \end{figure}

In contrast with the aligned population, nearly all surviving members of the anti-aligned population are initialized on \PX-crossing orbits. 
From Figure \ref{FIG:COP:acosw}, we see that this places the anti-aligned particles in a region of  phase-space suffused by an intricate mixture of regular and chaotic trajectories generated by the overlap of \PX~MMRs. 
Figure \ref{FIG:COPLANAR:WEBOUTER} illustrates some slices through this chaotic web in greater detail, showing two slices of phase space for particles with constant perihelion distances and semimajor axes both interior and exterior to Planet 9. 
In both panels, the initial conditions are such that they are anti-aligned with \PX's orbit (i.e., $\varpi=\varpi_9+\pi$).
In the top panel, corresponding to semimajor axes interior to \PX's orbit, \PX~is initially at pericenter and the test particle's initial mean anomaly is plotted on the vertical axes, which we have labeled $\phi$ to match notation introduced below in Section \ref{SECN:CHAOTICWEB:RESDYNAMICS}.
In the bottom panel, showing orbits exterior to \PX, the roles are reversed: the test particle is initially at pericenter and  \PX's initial mean anomaly is plotted on the vertical axes and labeled $\phi$.

We can understand much of the structure observed in Figure \ref{FIG:COPLANAR:WEBOUTER} by considering initial conditions that lead to a close encounter with \PX.
Let $M^{*}_{9,+}$ and $M^{*}_{9,-}$ be the mean anomalies of \PX~at the two points where the test particle's orbit intersects \PX's. Also, let $M^{*}_+$ and $M^{*}_-$ be the mean anomalies of the test particle at these same intersection points. 
The test particle will experience a close encounter if it reaches $M^{*}_\pm$ at the same time \PX~reaches $M^{*}_{9,\pm}$ after completing an integer number of orbits. 
In other words, there is a time $t$ that satisfies
\begin{eqnarray}
n t  &=& M^{*}_\pm + 2 \pi j \\
n_9 t &=& M^{*}_{9,\pm} + 2 \pi j_9
\end{eqnarray}
for some pair of integers $j$ and $j_9$ where $n$ and $n_9$ are the mean motions of the particle and \PX.
This condition is equivalent to 
\begin{eqnarray}
M^{*\pm}_{9,\pm} - \frac{n_9}{n} M^{*}_\pm =  2 \pi \frac{n_9}{n} j \mod{2\pi}~.
\label{EQ:CHAOTICWEB:COLLISION}
\end{eqnarray}
If the orbital frequency ratio ${n}/{n_9}$ is irrational, then the test particle will come arbitrarily close to \PX~after a sufficient amount of time has passed.
Close encounters can only be avoided if frequency ratio is of the form ${n}/{n_9}=k/N$ for integer $N$ and $k$, i.e. the particle is in resonance with \PX. 
In Figure \ref{FIG:COPLANAR:WEBOUTER}, we have traced the initial test particle orbital phases, as a function of semimajor axis, that lead to collisions with \PX~in less than two orbits going forward or backward in time.  
These lines trace clear features in the chaotic web of white pixels shown in both panels. 
By tracing initial conditions that lead to collisions after more and more orbits, we would fill in the ``backbone'' of the chaotic web. 
Continuing this tracing process indefinitely would densely fill the $a-\phi$ plane in Figure \ref{FIG:COPLANAR:WEBOUTER}, except at exactly resonant period ratios. 
Clearly, the entire phase-space is not densely filled by chaotic trajectories. 
This is because,  sufficiently close to a resonant period ratio, gravitational interactions cause the test particle's instantaneous orbital period to oscillate about the resonant value. 
Accordingly, this provides resonances with finite widths in semimajor axis space. 
A multitude of resonances are evident in Figure \ref{FIG:COPLANAR:WEBOUTER} as stable "islands" of black pixels in the white chaotic "sea."
The MMRs in  \figref{FIG:COPLANAR:WEBOUTER} exhibit obvious structure in the $a-\phi$ plane, which we describe in the next section. 
%%%%%%%%%%%%%%%%%%%%%%%%%%%%%%%%%%%%%%%%%%%%%%%%%%%%%%%%%%%%%%%%%%%%%%%%%
\subsection{Dynamics of a Single Resonance}
\label{SECN:CHAOTICWEB:RESDYNAMICS}
Before discussing chaos driven by overlap of MMRs, we first describe the dynamics of a single isolated MMR.
For concreteness, we will restrict our discussion to test particles in exterior resonances with \PX, though interior resonances can easily be considered by analogous means.
We will refer to MMRs occurring at period ratios $P/P_9=N/k$, where $N$ and $k$ are relatively prime integers, as $k$th order resonances. 
We caution the reader that this is {\it not} conventional terminology when referring to MMRs but, as \citet{Pan2004} note, it is an appropriate redefinition when considering highly eccentric orbits as we are here. 
While the dynamics of MMRs are often examined by considering the libration or circulation of the resonant angles 
\begin{eqnarray}
\phi=\frac{1}{k}\left(N\lambda-k\lambda_9 - (N-k)\varpi \right)\label{EQ:CHAOTICWEB:PHI}\\
\psi=\frac{1}{k}\left(N\lambda-k\lambda_9 - (N-k)\varpi_9\right)\label{EQ:CHAOTICWEB:PSI}
\end{eqnarray}
and their linear combinations, it is helpful to relate these angles to a more physical picture.
To this end, we will consider stroboscopic snapshots of the system at each pericenter passage of the test particle. 
When the test particle is at pericenter, $\lambda=\varpi$, and hence the resonant angle reduces to $\phi=(\varpi-\lambda_9)$, the angular separation between the test particle's longitude of pericenter and \PX's mean longitude. 

\begin{figure}
    \centering
    \includegraphics[width=0.45\textwidth]{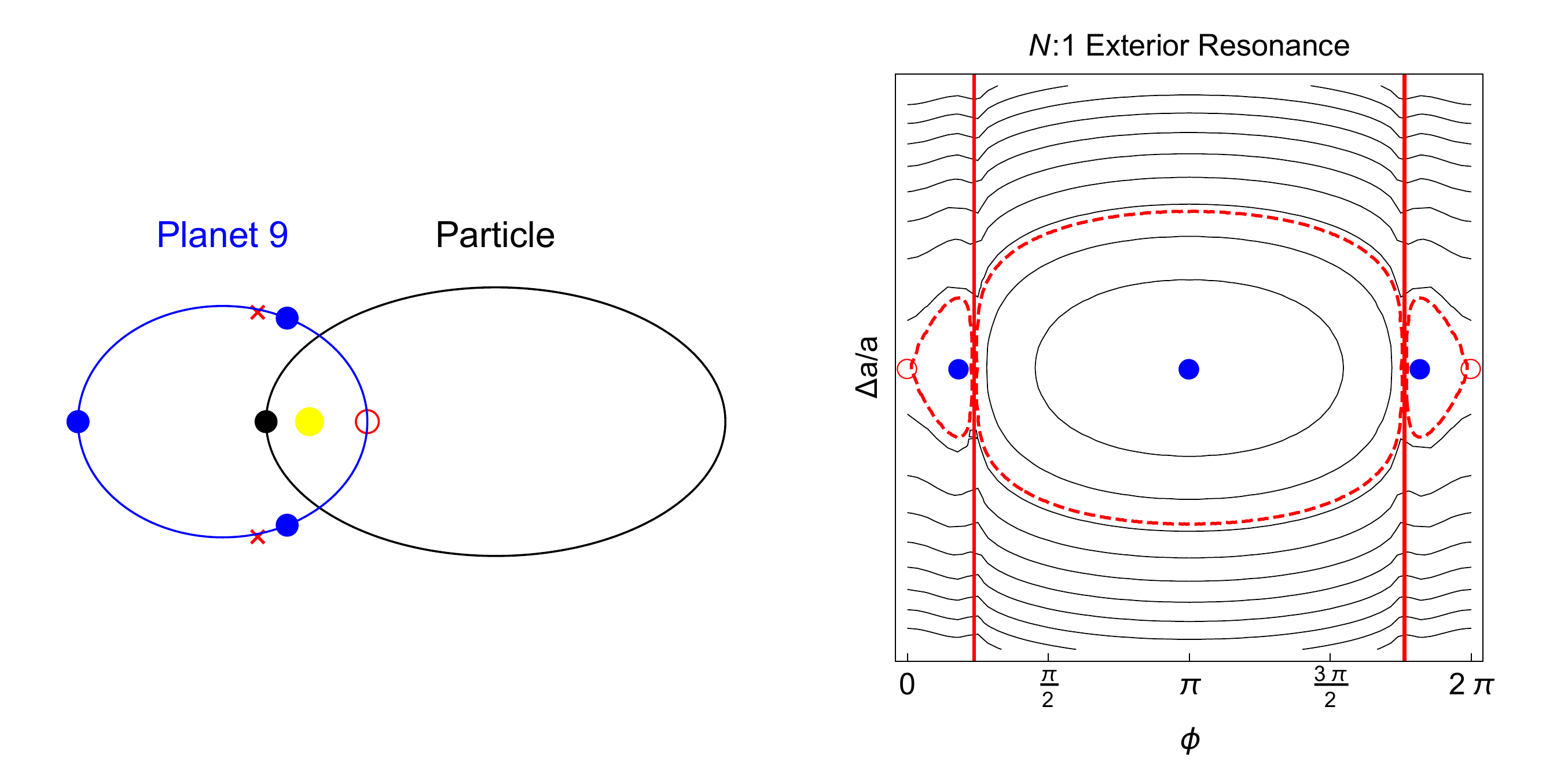}
    \includegraphics[width=0.45\textwidth]{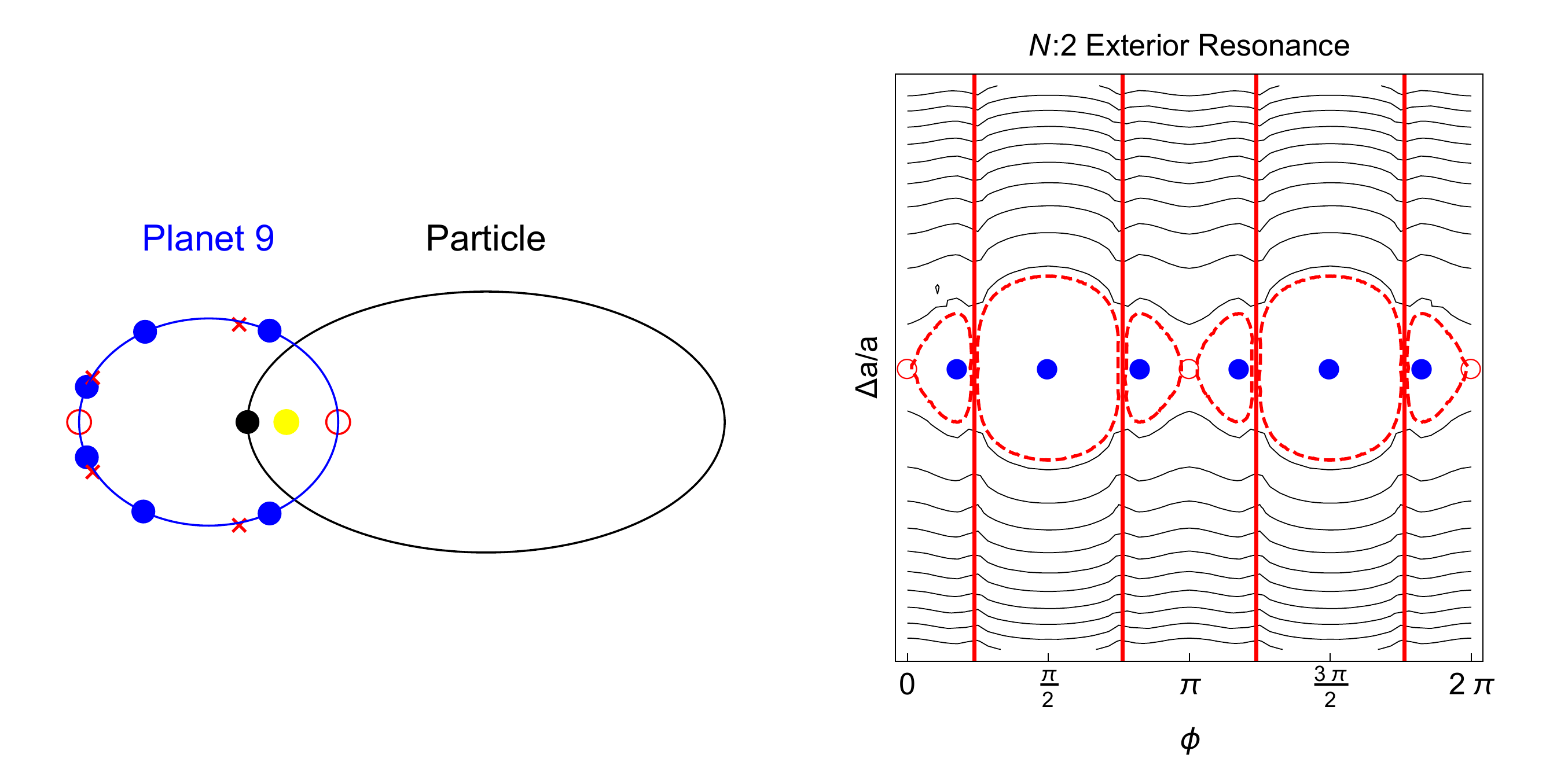}
    \caption{Schematic illustrations and phase-space portraits for test particles in $N$:1 and $N$:2 exterior resonances with \PX.
    Stable centers are indicated by blue points in the phase-space portraits and at the corresponding orbital phases in the schematics.
    Unstable fixed points are marked with red circles. The loci of points leading to collisions are shown as red lines in the phase space portraits and marked with red "x"s in the schematic.
    }
    \label{fig:res_portraits}
\end{figure}

Figure \ref{fig:res_portraits} shows representative phase-space portraits for  $N$:$1$ and $N$:$2$ exterior resonances with \PX~in the $\phi-\Delta a/a$ plane. The phase-space portraits show constant energy contours of resonant Hamiltonians constructed via an averaging procedure for particle orbits anti-aligned with \PX's (see Appendix \ref{SECN:APP:Equation} for details). 
The portraits are accompanied by schematic diagrams showing the location of \PX~in its orbit when the test particle is at pericenter for the various critical points of the phase space.
On timescales sufficiently shorter than the apsidal precession timescale, resonant test-particle trajectories are confined to the constant energy contours.\footnote{Only librating resonant trajectories are accurately captured by the averaged Hamiltonian. Circulating trajectories of the averaged resonant Hamiltonian will generally be chaotic trajectories of the full Hamiltonian from which the averaged resonant Hamiltonian is derived.}
The dashed red curves show the separatrix dividing librating and circulating trajectories.

First, we will consider a test particle in an exterior $N$:1 resonance with \PX.
Ignoring any orbital perturbations, after one complete test-particle orbit, \PX~will have returned to its initial orbital phase if the particle's orbital period is exactly commensurate with \PX's. 
In actuality, \PX's gravitational effects will change the test-particle's semimajor axis and cause apsidal precession.
Ignoring apsidal precession for the moment, the induced change in semimajor axis dictates the dynamics of the resonance on short timescales by altering the particle's orbital period.
An increase (decrease) in the test particle's orbital period causes a corresponding decrease (increase) in $\phi$ at the test particle's next pericenter passage, as \PX~will have advanced through slightly more (less) than $N$ complete orbits.
A fixed point of the resonance occurs when there is no net change to the test particle's semimajor axis over the course of its orbit, so $\phi$ is unchanged when the test-particle returns to pericenter.
For perfectly anti-aligned test particles orbits, $\phi=0$ and $\pi$ are always fixed points by symmetry. 
The point $\phi=\pi$, i.e., when \PX~is at apoapse during the particle's pericenter passage, is a stable fixed point: initial conditions that are slightly displaced from $(a,\phi)=(N^{2/3} a_9,\pi)$ where $a$ and $a_9$ are the semimajor axes of the particle and \PX, respectively, will oscillate about this point.
If these oscillations are too large, however, the test particle will collide with \PX~ at the point where the orbits cross. 
The loci of points corresponding to collisions with \PX~are shown in Figure \ref{fig:res_portraits} with red lines.
These collision points set the  maximum amplitude of stable oscillations in $\phi$ about the fixed point at $\pi$.
There are two additional stable `asymmetric libration' fixed points for $N$:$1$ resonances.\footnote{These asymmetric libration islands are reminiscent of the asymmetric libration islands studied by \citet{Beauge1994} for resonances in the circular restricted three-body problem. The islands are only stable if the indirect piece of the disturbing function is included in numerical averaging procedure, indicating that the Sun's reflex motion induced by \PX~plays an important dynamical role in the asymmetric libration islands.
} 
The $\phi$ location of the asymmetric libration points depends on the particle's perihelion distance.

The fixed points of an $N$:$k$ resonance with $k$>1, correspond to $k$-periodic orbits when viewed in terms of stroboscopic snapshots at pericenter: the test particle sees \PX~at a repeating series of $k$ distinct points at sequential pericenter passages. 
For example, in an $N:2$ MMR like the one shown at the bottom of Figure \ref{fig:res_portraits}, there is a stable fixed point associated with the test particle's pericenter passages occuring alternatively when \PX~is at $M_9=\pi/2$ then $M_9=3\pi/2$.
Additionally, there is an unstable fixed point corresponding to \PX's orbital phase occuring successively at $\phi=0,\pi$ as well as two pairs of $\phi$ values associated with asymmetric librations.

For interior resonances where the test particle's orbital period is less than \PX's, essentially every aspect of the dynamics described above are the same when one considers stroboscopic snapshots capturing \PX's pericenter passage rather than the test particle's. 

On a longer timescale, apsidal precession slowly modulates the phase-space portrait, changing the location of the resonance's fixed points and the shape of the resonant islands. Anti-aligned test particles in our fiducial numerical simulations do not precess far from perfect anti-alignment and so these effects are  modest.

%%%%%%%%%%%%%%%%%%%%%%%%%%%%%%%%%%%%%%%%%%%%%%%%%%%%%%%%%%%%%%%%%%%%
\subsection{Resonance overlap and the chaotic web}
\label{SECN:CHAOTICWEB:OVERLAP}
%%%%%%%%%%%%%%%%%%%%%%%%%%%%%%%%%%%%%%%%%%%%%%%%%%%%%%%%%%%%%%%%%%%%
Having examined the dynamics of individual resonances, we now turn to a more global picture of the dynamics of the anti-aligned particles.
In order to do so, we will approximate the dynamics of these particles by means of an area-preserving map.
This mapping captures the effects of the infinite number of MMRs with \PX~in any given semimajor axis interval and the chaos driven by their overlap.
The full derivation of our map, which we summarize here, is presented in Appendix \ref{SECN:APP:OTHER}.
The mapping approximates stroboscopic snapshots of a test particle's orbital energy $E$ and \PX's mean anomaly $\phi$ at each perihelion passage of the particle. 
Similar maps have been derived and used previously in the study of highly eccentric particles subject to a planetary perturber by a number of authors \citep[e.g.][]{PetroskyBroucke88,Malyshkin:1999,Pan2004,Shevchenko2011}.
Between each perihelion passage the gravitational influence of \PX~imparts a change in energy to the test particle.
Because the gravitational influence of \PX~is minimal when the particle is near aphelion and far from \PX, we approximate the particle's orbit as a parabolic orbit with the same perihelion distance as the true particle orbit.
The change to the particle's orbital energy changes its orbital period, which in turn affects \PX's orbital phase at the time of the particle's next perihelion passage.

\begin{figure*}[t]
    \centering
    \includegraphics[width=0.33\textwidth]{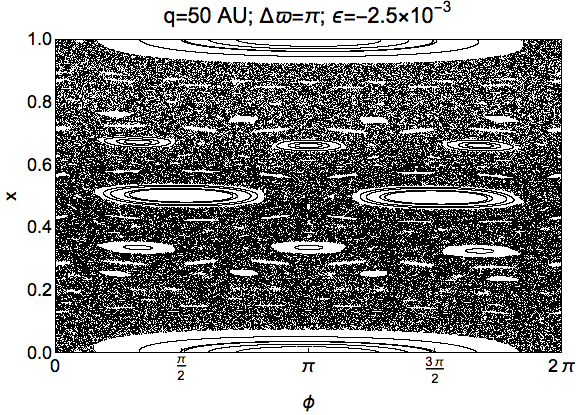}    
    \includegraphics[width=0.33\textwidth]{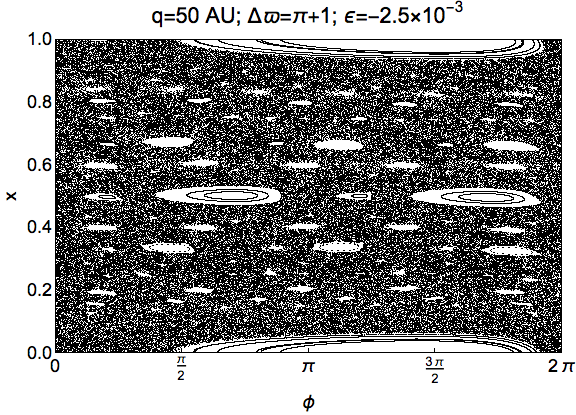}    
    \includegraphics[width=0.33\textwidth]{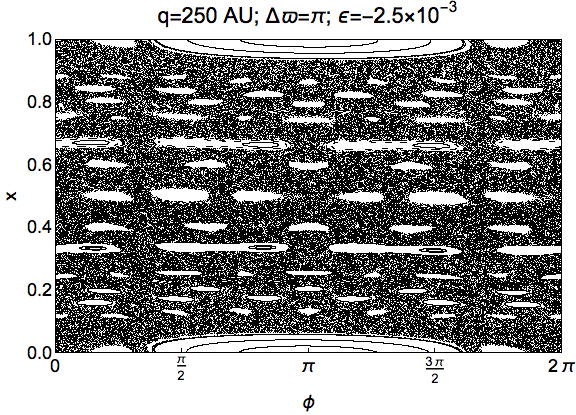}
    %%%
    \includegraphics[width=0.33\textwidth]{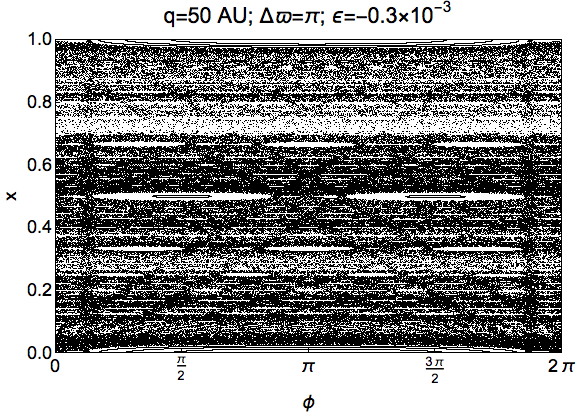}     
    \includegraphics[width=0.33\textwidth]{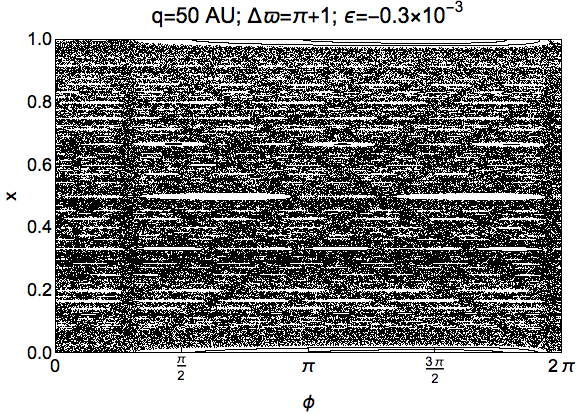} 
    \includegraphics[width=0.33\textwidth]{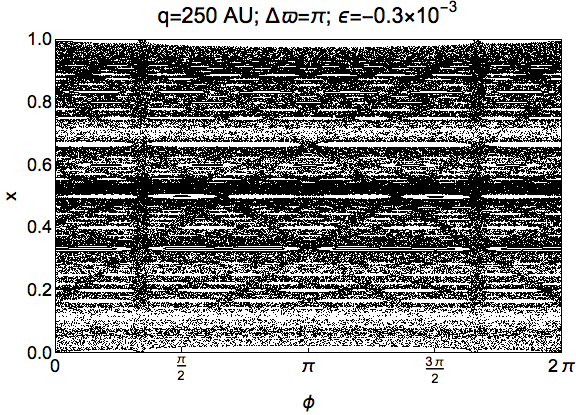}
    \caption{Some example phase-space portraits of the mapping of Equation \ref{Equation:Map2} for different parameters. 
    The top and bottom rows show maps made with the same $q$ and $\Delta\varpi$ for two different values of $\epsilon=-(3/2)\mu_9 \fracbrac{a_\text{res}}{ a_9 }^{5/2}$: the top row corresponds to \emph{higher} values of \PX's mass and/or particle semimajor axis.
    The $x$ coordinate is plotted modulo 1. See text for more details.}
    \label{FIG:CHAOTICWEB:P9MAPPINGS}
\end{figure*}

The mapping  is given by
\begin{eqnarray}
E'    &=& E + \mu_9 \delta E(\phi; q,\Delta\varpi) \nonumber 
\\
\phi' &=& \phi + 2\pi \fracbrac{E_0}{E'}^{3/2} \mod{2\pi}\label{Equation:Map1}
\end{eqnarray}
where primed and un-primed variables represent values before and after a mapping step, respectively, $\delta E$ is a function that gives the increment in $E$ as a function of \PX's orbital phase, as well as the test-particle perihelion distance $q$ and apsidal alignment $\Delta\varpi=\varpi-\varpi_9$, and $E_0$ is the orbital energy of a test particle with the same semimajor axis as \PX.
We describe our method for determining $\delta E$ via numerical integration in Appendix \ref{SECN:APP:Equation}.
Because particles' eccentricities and orbital alignments evolve slowly, compared to their semimajor axes, we treat $q$ and $\Delta\varpi$ as fixed parameters when applying the mapping.

To get a better understanding of the local phase space structure in a given semimajor axis range, we expand the mapping, Equation \ref{Equation:Map1}, about a semimajor axis, $a_\text{res}$, corresponding to an $N$:$1$ MMR with \PX.
We define a new variable $x=2N(a-a_\text{res})/3a_\text{res}$ and obtain the new mapping
\begin{eqnarray}
x' &=& x + \epsilon \delta E(\phi; q,\Delta\varpi) \nonumber 
\\
\phi' &=& \phi - 2\pi x' \mod{2\pi}\label{Equation:Map2}
\end{eqnarray}
where $\epsilon=-(3/2)\mu_9 \fracbrac{a_\text{res}}{ a_9 }^{5/2}$.
% reflects the fact that
Particles become more loosely bound as the semimajor axis, $a_\text{res}$, increases, such that  a smaller $\mu_9$ can have the same dynamical effect at large $a_\text{res}$ as a greater $\mu_9$ does at a smaller  $a_\text{res}$.
This is reflected in $\epsilon$'s scaling with $\mu_9$ and  $a_\text{res}$.
In the new mapping, Equation \ref{Equation:Map2}, first-order resonances occur at integer values of $x$  where $\phi$ advances by an integer multiple of $2\pi$.
The mapping in Equation \ref{Equation:Map2} is unchanged if we add an integer to $x$ so we can obtain a complete picture of the phase-space structure by taking $x$ modulo 1.
Therefore, phase-space regions between successive $N$:$1$ MMRs are identical to the extent that the linearization approximation we used to derive Equation \ref{Equation:Map2} remains valid.
Repeating, nearly-identical phase-space structure between successive $N$:$1$ MMRs is evident in the stability map shown in the bottom panel of Figure \ref{FIG:COPLANAR:WEBOUTER}.
Equation \ref{Equation:Map2} provides a good approximation of a test particle's dynamics so long as $|x|\ll N$ since we assumed that $|a-a_\text{res}| \ll a_{res}$ to derive it from  Equation \ref{Equation:Map1}.

Some sample mapping trajectories of Equation \ref{Equation:Map2} are plotted in Figure \ref{FIG:CHAOTICWEB:P9MAPPINGS}. 
These trajectories reveal a number of notable features of the dynamics of the anti-aligned population:
\begin{enumerate}
\item Chaotic trajectories fill much of the space for all parameter values.
There are no KAM curves spanning the full phase space from $\phi=0$ to $\phi=2\pi$.
Therefore, there should be no barriers in phase space bounding the chaotic diffusion of test particles' semimajor axes. The lack of bounding KAM curves reflects the fact that, for any initial period ratio, there are initial values of $\phi$ that result immediately in a close encounter with \PX.
\item Increasing $\epsilon$  increases the size of low-order resonant islands while subsuming many of the higher-order resonant islands in the chaotic sea.
\item The middle panels of Figure \ref{FIG:CHAOTICWEB:P9MAPPINGS} demonstrate the effects of a modest deviation from exact anti-alignment:
the resonance centers become shifted in $\phi$ and the resonant islands become slightly distorted in shape. Otherwise, much of the resonant structure remains unchanged.
\item Comparing the left- and right-hand columns of Figure \ref{FIG:CHAOTICWEB:P9MAPPINGS} illustrates the influence of the  perihelion distance, $q$, on the test-particle dynamics. Here,  $q$ determines the extent of resonant islands in the $\phi$ coordinate by setting the $\phi$ value at which collisions occur.
This same effect was seen to determine the extent of individual resonances in Section \ref{SECN:CHAOTICWEB:RESDYNAMICS}.
\end{enumerate}
% Clearly resonances become more overlapped for larger values of $\epsilon$ and there are fewer regular trajectories around resonant islands.
The lack of bounding KAM curves for any set of parameters mapping Equation \eqref{Equation:Map2} is atypical, compared to the behavior of traditional perturbed twist mappings such as the standard map.
The map lacks bounding KAM curves  because the  perturbation,  $\delta E$, is discontinuous at $\phi$ values corresponding to collision, violating smoothness conditions necessary for such curves to exist \citep[e.g.][]{LLBook}.\footnote{Given this feature of our mapping, its not entirely clear to what degree all the observed chaos should be appropriately ascribed to resonance overlap versus the unsmooth orbital evolution introduced by close encounters with \PX.}
A mutual inclination between \PX~and would smooth out the discontinuities $\delta E$, possibly allowing for KAM curves to appear. 
We return to this point below in our discussion in Section \ref{SECN:DISC}.
%

%%%%%%%%%%%%%%%%%%%%%%%%%%%%%%%%%%%%%%%%%%%%%%%%%%%%%%%%%%%%%%%%%%%%%%%%

\subsection{The role of Neptune}
\label{SECN:CHAOTICWEB:NEPTUNE}

\begin{figure*}
    \centering
     \includegraphics[width=0.33\textwidth]{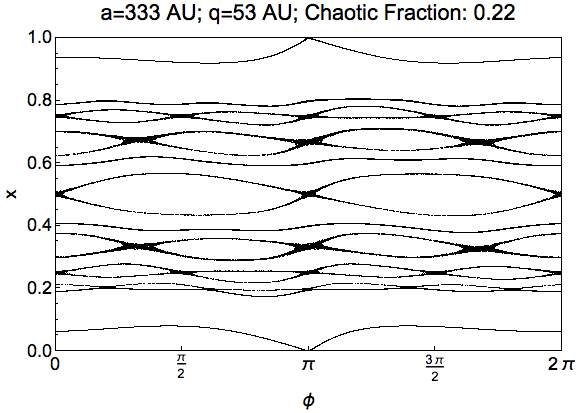}
     \includegraphics[width=0.33\textwidth]{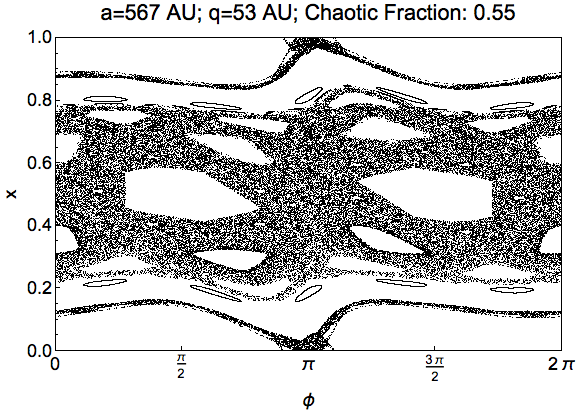}
     \includegraphics[width=0.33\textwidth]{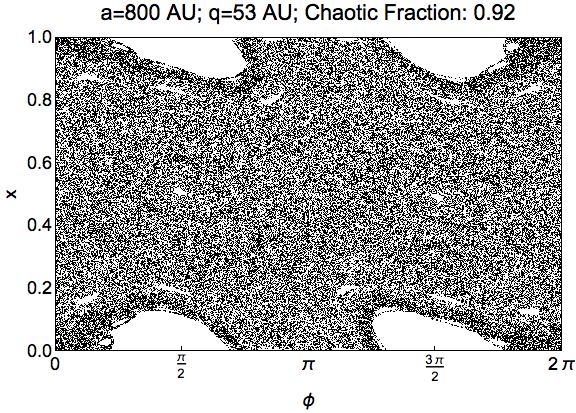}
    \caption{
    Three examples of increasing chaotic fraction of trajectories of particles perturbed by Neptune using the mapping Equation \eqref{Equation:Map2} for three different semimajor axis values. 
    In each panel, trajectories are started at $(x,\phi)=(x_0,\pi)$ where $x_0=p/q$ is a rational number with $q\le 5$.
    These initial conditions place the trajectories near unstable fixed points of resonances up to fifth order.
    The trajectories explore the extent of the chaotic region filled by the separatrix layers of each resonance. 
    }
\label{fig:neptune_maps}
\end{figure*}

\begin{figure}
    \centering
     \includegraphics[width=0.45\textwidth]{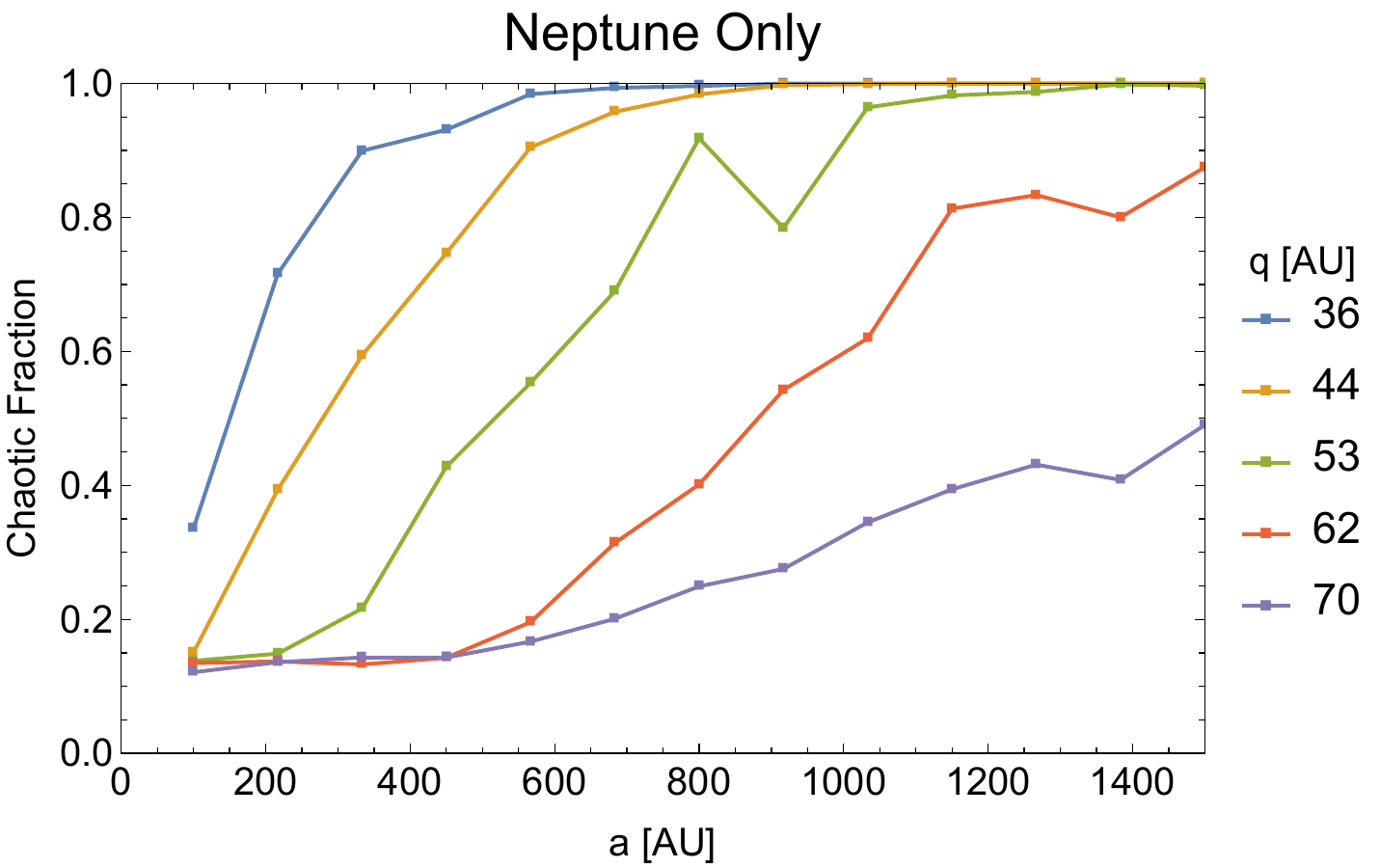}
    \caption{
    Chaos fraction computed with mapping Equation \ref{Equation:Map2} for particles subject to perturbations by Neptune as a function of semimajor axis for different perihelion distances.
    %The chaos fraction is computed by dividing the mapping's phase space into a 100$\times$100 grid and counting the fraction grid cells that one or more trajectory point.
    A significant portion of the initial conditions in our fiducial simulation ($q\in[33,50]\au$, $a\in[150,550]\au$) occupy regions of phase space where the motion is expected to be chaotic under the influence of Neptune alone.
    }
    \label{fig:neptune_chaos_frac}
\end{figure}

We saw in Section \ref{SECN:NUMERICAL:OVERVIEW} that including an active Neptune in numerical simulations leads to significantly more chaotic test-particle behavior. 
When Neptune's influence is approximated only by its quadrupole contribution to the gravitational potential, most of the long-lived test-particles librate stably in MMRs with \PX. \citet{BatyginMorbidelli17} find similar stable, librating behavior in their simulations that approximate the effect of all of the giant planets by a quadrupole potential.
By contrast, when the full effect of Neptune is taken into account, most of the test particles' semimajor axes diffuse chaotically.

We can apply the same mapping derived above in Equation \eqref{Equation:Map2} in order to assess Neptune's contribution to particles' chaotic semimajor axis evolution.
Figure \ref{fig:neptune_maps} show some example chaotic trajectories of particles perturbed solely by Neptune near three different semimajor axis values. 
The trajectories were computed using Equation \ref{Equation:Map2}, but with parameters appropriate to Neptune.\footnote{Specifically, we set $\epsilon=(-3/2)\mu_N(a/a_N)^{5/2}$ where $\mu_N=5.15\times10^{-5}$ is Neptune's mass ratio relative to the Sun and $a_N=30\au$ is Neptune's semimajor axis.
We assume a circular orbit for Neptune, so $\delta E$ is independent of $\Delta\varpi$.}
Each panel shows trajectories of points started near the unstable fixed points of  resonances up to fifth order.
We compute the `chaotic fraction' of the mapping's phase space by dividing the phase space into a 100$\times$100 grid and counting the fraction of grid cells that contain at least one point of any trajectory. Given the fractal nature of the chaotic phase space, the measured fraction is sensitive to the size of grid cell used. Nonetheless, the computed chaotic fractions illustrate that the full phase-space of the mapping becomes progressively more chaotic with increasing semimajor axis, corresponding to larger values of  $\epsilon$.
This point is further emphasized in Figure \ref{fig:neptune_chaos_frac}, where we plot chaos fractions computed for mapping Equation \ref{Equation:Map2} as a function of semimajor axis for different perihelion distances.
Clearly, a significant portion of the initial conditions in our fiducial simulation ($q\in[33,50]\au$, $a\in[150,550]\au$) occupy regions of phase space where the motion is expected to be chaotic under the influence of Neptune alone.

%%%%%%%%%%%%%%%%%%%%%%%%%%%%%%%%%%%%%%%%%%%%%%%%%%%%%%%%%%%%%%%%%%%%%%%%%
%%%%%%%%%%%%%%%%%%%%%%%%%%%%%%%%%%%%%%%%%%%%%%%%%%%%%%%%%%%%%%%%%%%%%%%%
\subsection{Eccentricity and apsidal evolution}
\label{SECN:CHAOTICWEB:APSIDAL}
\begin{figure}
     \centering
     \includegraphics[width=\columnwidth]{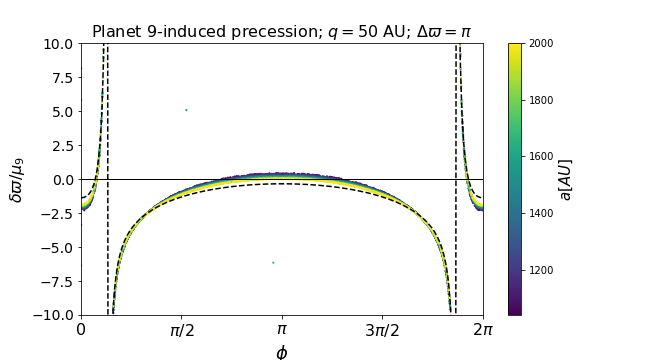}
     \caption{Perihelion precession induced by \PX~in a single particle perihelion passage. Points show $N$-body results for particles, colored according to semimajor axis and spanning a range $a\in[1000,2000]\au$. An analytic prediction, derived by approximating particle orbits as parabolic, is shown by the black dashed line. The perihelion precession induced per orbit is nearly independent of a particle's semimajor axis and well-approximated by the analytic prediction.}
     \label{FIG:CHAOTICWEB:P9PRECESSION}
 \end{figure}

 Gravitational interactions with Neptune and \PX~affect test particles' angular momentum, as well as their energy.  Therefore, test particles' eccentricities and longitudes of pericenter should evolve at the same time as they experience chaotic semimajor axis evolution.  
 In the numerical integrations presented in Section \ref{SECN:NUMERICAL},  we saw that the chaotically scattering population maintains orbital anti-alignment with \PX~to within roughly $\sim \pm 60~\deg$. 
 In Section \ref{SECN:SECULAR}, we showed that, in the secular approximation, a stable anti-aligned libration island exists where torques from \PX~and Neptune's averaged gravitational potentials balance to maintain anti-aligned particle orbits. 
 However, these particles' chaotic semimajor evolution invalidates the usual assumptions made when employing a secular approach.
 Nonetheless, we argue that the evolution of test particles' eccentricities and longitudes of perihelion should, on average, follow the predictions of secular theory. 
 In part, this is because, at high eccentricity, the torques causing changes in angular momentum and longitude of perihelion vary little with semimajor axis. 
 Consequently, the particles' chaotic semimajor axis diffusion has little effect on their angular momentum dynamics. 
 The torques are largely independent of semimajor axis because the gravitational effects of Neptune and \PX~are concentrated near pericenter when interactions are strongest. 
 In particular, the average precession rate induced by Neptune reduces to 
 \begin{equation}
 \dot{\varpi} =\frac{3}{4}n\mu_N \fracbrac{a_N}{a(1-e^2)}^2 \approx \frac{3}{4}n \mu_N \fracbrac{a_N}{2q}^2 \label{EQ:CHAOTICWEB:J2} \end{equation}
 for large $a$ at fixed $q$. 
 According to Equation \eqref{EQ:CHAOTICWEB:J2}, the average perihelion precession induced  by Neptune {\it per orbit} depends only on $q$.
 Similarly, the per-orbit precession induced by \PX~ depends only on $q$ for a sufficiently large particle $a$. 
 Figure \ref{FIG:CHAOTICWEB:P9PRECESSION} compares the increment in $\varpi$ induced by \PX~after a single orbit, as measured in $N$-body simulations of particles with semimajor axes spanning the range $a\in[1000,2000]\au$.
 An analytic prediction that approximates the test-particle orbit as parabolic is also shown (see Appendix \ref{SECN:APP:OTHER} for details). 
 The analytic prediction depends only on the perihelion distance, $q$, and provides an excellent fit to the $N$-body simulation results. 
 
The near-independence of torques on semimajor axis alone does not fully justify the secular approximation as an explanation for why anti-alignment is maintained.
In the usual secular averaging procedure, one performs a double integral over the orbital phases of the test particle and \PX.  
The validity of this double averaging requires that the orbital phases of the particle and \PX~are uncorrelated, e.g., there are no MMRs between \PX~and the test particle.
The mapping shows that the uncorrelated assumption is valid for the chaotically diffusing particles as well.
It is straightforward to show, by making a histogram of the mapping variable $\phi$ for some chaotic trajectories of the mapping Equation \eqref{Equation:Map1}, that  
the distribution of $\phi$ for these trajectories is nearly uniform, demonstrating that \PX's orbital phase is not correlated with the test particles' phases.
Therefore, the double-integral over orbital phase used in the usual secular approximation is equally valid in the case of the chaotically diffusing particles.
\begin{figure}
    \centering
    \includegraphics[width=\columnwidth]{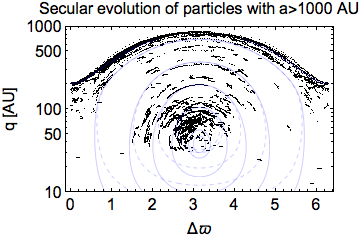}   
    \caption{Secular evolution of test particles with large semimajor axes.
    The black points show $N$-body results from our fiducial simulation. 
    Blue lines show contours of secular Hamiltonians computed with $a=1500$\au~(dashed) and $a=3000$\au~(solid).
    The precise choice of semimajor axis has a modest effect on the contour shapes, as described in the text.}
    \label{FIG:CHAOTICWEB:SECULARLARGEA}
\end{figure}

In \figref{FIG:CHAOTICWEB:SECULARLARGEA}, we plot perihelion distance, $q$, versus orbital alignment $\Delta\varpi$ for particles with $a>1000\au$ from our fiducial simulation. 
We also plot contours of secular Hamiltonians computed for  $a=1500$ and $3000\au$. These contours are produced in essentially the same way as the secular contours plotted in Figure \ref{SECN:SECULAR}  \citep[see also][]{BEUST16,BatyginMorbidelli17}, except now we plot $q$ versus $\Delta\varpi$ instead of $e\cos\Delta\varpi$ and $e\sin\Delta\varpi$.
As argued above, the shape of these contours is insensitive to the choice of $a$ when $a$ is large, so these contours should be representative of  the secular trajectories followed by all of the particles plotted in \figref{FIG:CHAOTICWEB:SECULARLARGEA}. Indeed, we see that the tracks followed by particles in the $q-\Delta\varpi$ plane closely follow the predicted secular contours. This demonstrates that secular theory can be used to successfully predict the apsidal evolution of these particles despite their significant diffusion in semimajor axis.

The low-$q$ points in Figure \ref{FIG:CHAOTICWEB:SECULARLARGEA} are not uniformly spaced on their secular trajectories, but rather cluster at $q$ values above the equilibrium point located near $(q,\Delta\varpi)=(\sim35\au,\pi)$.
We observe in our simulations that when particles' perihelion distances dip below $q=30\au$, they are quickly ejected by Neptune.
Eventually, most of the low-$q$ particles shown in Figure \ref{FIG:CHAOTICWEB:SECULARLARGEA} should evolve onto Neptune-crossing orbits and be removed. However, the timescale to do so is longer than our simulation duration and we are left only with particles whose secular evolution has stalled at perihelion distances $q>30\au$, causing  the observed clustering.

\begin{figure*}
    \centering
    \includegraphics[width=\columnwidth]{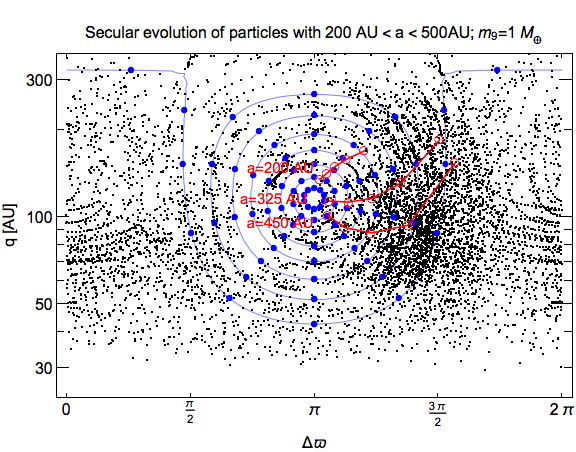}
     \includegraphics[width=\columnwidth]{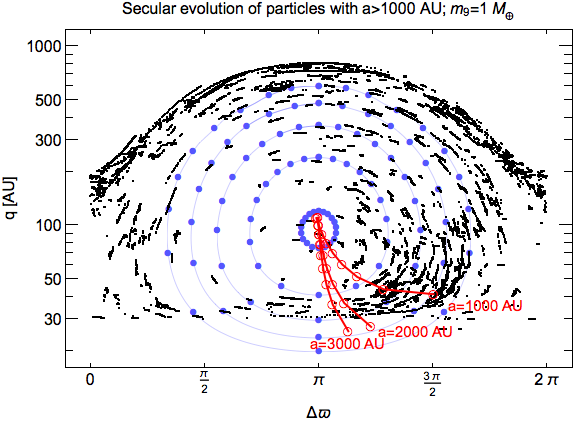}
    \caption{
    Secular evolution of test particles subject to perturbations by a $1\Mearth$ \PX.
    Black points show $N$-body results from our low-mass simulation.
    {Left panel:} 
    particles interior to \PX~with $200\au<a<500\au$. 
    Blue lines show contours of the secular Hamiltonian for $a=325\au$.
    The red lines show the endpoints reached by secular trajectories initialized  with $\Delta\varpi=\pi$ at three different semimajor axes, $a=200, 325$ and $450\au$, after 5 Gyrs. 
    The initial $q$ values of these trajectories are chosen to span from the stable equilibrium point at each $a$ value to the minimum $q$ value contained within the librating region.
    A series of blue points uniformly spaced in time are plotted on each contour.
    {Right  panel:} 
    particles exterior to \PX~with $a>1000\au$.
    The blue lines show contours of the secular Hamiltonian with $a=3000\au$.
    The red lines show the endpoints reached by secular trajectories at three different semimajor axes, $a=1000,2000$, and $3000\au$, after 5 Gyrs.
    The secular trajectories are initialized  with  $\Delta\varpi=\pi$ and $q$s that span from $30\au$ to the equilibrium point.
    }
    \label{FIG:CHAOTICWEB:SECULARLARGEA_LOWMASS}
\end{figure*}

In Section \ref{SEC:OTHERMASSES} we saw that when  \PX's mass was reduced to  $1\Mearth$, the chaotically scattering particles clustered near $\Delta\varpi=3\pi/2$ (rather than $\Delta\varpi=\pi$).
We argue that this clustering reflects the fact that the particles have not had sufficient time to relax from their initial conditions. Figure \ref{FIG:CHAOTICWEB:SECULARLARGEA_LOWMASS} plots $q$ versus $\Delta\varpi$ for particles both interior and exterior to \PX~in the $1\Mearth$ \PX~simulation. Representative secular Hamiltonian contours are plotted as well.  In order to demonstrate that secular evolution has not had sufficient time to randomize particle alignments, we integrate the equations of motion of the secularly averaged Hamiltonian at some representative values of particle semimajor axes. At a given semimajor axis, we initialize a set of secular trajectories on contours contained within the anti-aligned libration region. These trajectories are initialized at their minimum $q$ values, which occur at $\Delta\varpi=\pi$. The trajectories' evolution proceeds counter-clockwise about the secular contours, and we plot the final $q$ and $\Delta\varpi$ after 5 Gyr as red lines in Figure \ref{FIG:CHAOTICWEB:SECULARLARGEA_LOWMASS}. In the left panel of that figure, we see that the secular trajectories of particles interior to \PX~complete approximately $\sim1/4$ of a precession cycle over the course of 5Gyr, leaving them with $\Delta\varpi\sim 3\pi/2$.\footnote{{There is nothing dynamically significant about the exact value $\Delta\varpi=3\pi/2$. Rather, $\Delta\varpi\sim 3\pi/2$ is just an approximate average of the $\Delta\varpi$ values reached by the secular trajectories plotted in Figure \ref{FIG:CHAOTICWEB:SECULARLARGEA_LOWMASS} after 5 Gyrs.}}  Because at least one secular precession cycle must occur for initial $\Delta \varpi$ values to be effectively randomized, at least $\gtrsim 20$ Gyr of evolution would be neccessary for particles to relax to an equilibrium distribution. Trajectories exterior to \PX~shown in the right panel experience an  even smaller fraction of their secular precession cycle. While the apsidal evolution experienced by the $N$-body particles plotted in Figure \ref{FIG:CHAOTICWEB:SECULARLARGEA_LOWMASS} is more complicated than these simple secular trajectories because of semimajor axis diffusion, the $N$-body particles nonetheless fall within a region of the $(q,\Delta\varpi)$ plane similar to the secular trajectories.

The clustering is not just an artifact of initializing test particles in exactly (anti-)aligned configurations.
We observe a similar clustering in simulations initialized from randomized initial alignments, $\varpi$.
In these randomized simulations, particles precess relatively quickly through the bottom portion of  the secular trajectories plotted in \figref{FIG:CHAOTICWEB:SECULARLARGEA_LOWMASS}, where their precession is driven by mainly by Neptune. Their evolution slows dramatically upon reaching $\Delta\varpi\sim3\pi/2$ where their $q$s begin to increase and, consequently, the rate of Neptune-induced precession decreases (Equation \eqref{EQ:CHAOTICWEB:J2}).
Thus, even though the particles are initially randomized in $\varpi$, they have not relaxed from their initial narrow range of $q$s, which place them near the bottom of the secular trajectories plotted in Figure \ref{FIG:CHAOTICWEB:SECULARLARGEA_LOWMASS}.

Our simulations show that a $1\Mearth$ \PX~is capable of inducing clustering among ESDOs via a dynamical mechanism very different from the one operating in our simulations with a $10\Mearth$ \PX.
Taken at face value, this poses a serious impediment to deriving indirect dynamical constraints on \PX's mass and orbit because there is no obvious way to observationally distinguish which mechanism is responsible for the ESDOs' apsidal clustering. 
However, it remains to be seen whether the $1\Mearth$ \PX~clustering mechanism operates effectively in more realistic simulations allowing for mutual inclinations.
The results of \BBb, who disfavor such a small~\PX~mass based on an extensive grid of $N$-body simulations, suggest that inclinations probably render the $1\Mearth$ \PX~clustering mechanism less effective.

%%%%%%%%%%%%%%%%%%%%%\P%%%%%%%%%%%%%%%%%%%%%%%%%%%%%%%%%%%%%%%%%%%%%%%%%%%%
%%%%%%%%%%%%%%%%%%%%%%%%%%%%%%%%%%%%%%%%%%%%%%%%%%%%%%%%%%%%%%%%%%%%%%%%%
\section{Discussion}
\label{SECN:DISC}
The goal of this work has been to identify the dominant dynamical mechanisms governing the orbital evolution of TNOs in the presence of \PX.
Specifically, we set out to find the dynamical explanations for the features observed in our numerical simulations and enumerated in Section \ref{SEC:SIMSUMMARY}.
We can now address the dynamical origin of each of these features:
\begin{enumerate}
\item{
We identified an ``aligned" population of surviving particles inside  $a_\text{crit}\sim 250\au$ that circulate or maintain alignment with \PX's orbit while experiencing little semimajor evolution.
We showed in Section \ref{SECN:SECULAR} that the dynamical behavior of this population is well-explained by a standard secular approximation of the dynamics. Furthermore, secular evolution sets the maximum semimajor axis, $a_\text{crit}$, to which the aligned population extends. Initially, \PX-aligned particles beyond $a_\text{crit}$ secularly evolve onto (nearly) \PX-crossing orbits, resulting in their ejection.}
\item{ 
We noted that, beyond $a_\text{crit}$, most surviving particles from the fiducial simulation are anti-aligned with \PX's orbit and maintain anti-alignment over the course of their orbit.
While the secular approximation is not strictly valid for these particles because they experience significant chaotic semimajor axis evolution, their persistent anti-alignment qualitatively agrees with the presence of a stable, anti-aligned libration island in the secular Hamiltonian presented in Section \ref{SECN:SECULAR}.
Despite the fact that the usual assumptions of the secular approximation are not valid for such particles, we showed in Section \ref{SECN:CHAOTICWEB:APSIDAL} that their apsidal evolution should follow contours of a secular Hamiltonian. For large sufficiently large $a$, the amount of perihelion precession per orbit depends only on the perihelion distance, so these contours do not depend on $a$.
}
\item{
We noted that essentially no surviving particles transitioned from aligned to anti-aligned configurations or vice versa.  Such particles would have had to evolve through an orbit that meets \PX's tangentially near apoapse. As discussed in  Section \ref{SECN:CHAOTICWEB:ALIGNED}, this causes strong resonance overlap that results in the ejection  of such particles.
}
\item{
Particles in the fiducial simulation are occasionally captured in resonances, but do not remain stably librating in resonance for the full duration of the simulation. 
In Section \ref{SECN:CHAOTICWEB:OVERLAP}, we described the phase space structure explored by the chaotically scattering anti-aligned populations.
The phase space is filled everywhere with a mixture of resonant islands, along with chaos driven by encounter trajectories and the overlap of resonances.
Particles explore the chaotic region of this phase space. 
Resonance sticking can occur when chaotically diffusing particles come near a resonant island.
This sticking behavior is a generic feature of area-preserving twist maps \citep{LLBook}.
In Section \ref{SECN:CHAOTICWEB:NEPTUNE} we showed that Neptune can induce significant chaotic behavior in many of the TNOs in our simulations. This accounts for the significant difference between anti-aligned particles in our fiducial simulation and the quadrupole simulation.
}
\item{
The distribution of anti-aligned particles' $\Delta\varpi$ in the 1$M_\earth$ simulations, clustering near $3\pi/2$, can be explained by the fact that it has not relaxed to an equilibrium distribution because the secular precession timescale is longer than the age of the solar system. 
% duration of our simulations.
}
\end{enumerate}

We have relied on simplified coplanar simulations to help isolate and identify individual dynamical effects.
The dynamical mechanisms identified in this work can help us better understand the results of more realistic numerical simulations and ultimately help to constrain the mass and orbital properties of a putative \PX~based on observational data.

We have ignored the gravitational influence of the other solar system giant planets. Jupiter, Saturn, and Uranus combined induce additional perihelion precession at a rate approximately equal to that induced by Neptune.
\citet{BEUST16} and \citet{BatyginMorbidelli17} model the secular behavior of particles by accounting  for the effect of  \PX~and the quadrupole potential of the complete outer solar system.
The qualitative features of particles' secular evolution remain unchanged: apsidally anti-aligned libration islands are present at high eccentricity on crossing orbits with \PX.
The other giant planets' contribution to particle semimajor axis evolution is expected to be negligible compared to Neptune's. 
This is easily demonstrated by applying the  mapping method used in Section \ref{SECN:CHAOTICWEB:OVERLAP} and \ref{SECN:CHAOTICWEB:NEPTUNE} to model the influence of the other outer solar system planets.

Most anti-aligned particles in our fiducial simulation have semimajor axes exterior to \PX's. 
Consequently, in Section \ref{SECN:CHAOTICWEB} we focused on the dynamics of anti-aligned exterior particles.
It is straightforward to adapt the mapping model in Section \ref{SECN:CHAOTICWEB:OVERLAP} to describe the evolution of particles interior to \PX~by simply considering stroboscopic snapshots at \PX's perihelion passage rather than the particle's.
Many of the qualitative features of the dynamics remain the same as can be seen in the top panel of \figref{FIG:COPLANAR:WEBOUTER}:
the phase space consists of regular resonant islands embedded in a chaotic sea caused by close encounters and resonance overlap.
The structure of resonance locations repeats between successive $N:1$ resonances. (These resonances become progressively more tightly spaced toward smaller semimajor axes at the left of the plot.) 
While we find that most anti-aligned particles orbit exterior to \PX, the observational sample of clustered TNOs that motivate the \PX~hypothesis are mainly interior to \PX's orbit.
On its face, this would suggest that our simulations do not support the \PX~hypothesis.
However, TNOs on orbits exterior to \PX~spend a larger fraction of their orbit at distances where they are unobservable. 
Therefore, there is a strong observational bias toward TNOs on orbits interior to \PX~even if there are significantly more on orbits exterior to \PX.  
Additionally, relaxing our coplanar assumption may reduce \PX's disruptive effect on  particles initialized interior to it by allowing particles to avoid close encounters.
%

%We have only considered coplanar particles and planets in this paper.
%
Even modest mutual inclinations between TNOs and \PX~can yield a rich range of additional dynamical behaviors not captured by coplanar simulations. 
Three-dimensional simulations produce a chaotic anti-aligned population similar to that seen in our fiducial simulation.
This population maintains relatively modest mutual inclinations ($i\lesssim40\deg$) relative to \PX.
We expect the chaotic phase space explored by these particles to be qualitatively similar to the one we have described in \ref{SECN:CHAOTICWEB}: the mapping model used in Section \ref{SECN:CHAOTICWEB:OVERLAP} can easily be adapted to treat particle orbits inclined relative to \PX's. 

The mutually inclined case does present one important difference from the coplanar case:  inclined particles can avoid close encounters with \PX~without the need for the phase protection offered by MMRs.
Mutual inclination smooths out the discontinuity in $\delta E(\phi;q,\Delta\varpi)$ in Equation \eqref{Equation:Map1} that occurs where the test-particle trajectory collides with \PX.
For sufficiently large inclinations, $\delta E(\phi;q,\Delta\varpi)$ may become smooth enough to allow KAM curves that bound regions of the mapping's phase space and limit semimajor axis diffusion.
As particle inclinations evolve, these curves could appear and disappear as the whole phase space is slowly  modulated by the mutual inclination between the particle and \PX.
We have already observed an analogous scenario in Section  \ref{SECN:CHAOTICWEB:ALIGNED}, where the slow secular evolution of initially aligned particles brings them in and out of a region of strong resonance overlap near orbit-crossing with \PX.
\citet{BatyginMorbidelli17} suggest a similar scenario to explain the dynamical behavior observed in their simulations of inclined test particles. They postulate that inclination evolution takes particles into and out of regimes where close encounters with \PX~drive chaotic semimajor axis evolution. 
They posit that, in between periods of stochastic evolution, particles settle into libration in MMRs with \PX.
The assertion that chaotic evolution is punctuated by periods of {\it resonant libration} seems unwarranted: particles could just as well find themselves in regions of bounded chaos or on KAM curves with irrational winding numbers as inclinations evolve out of regimes of strong chaos driven by close encounters.
We plan to explore the effects of mutual inclinations in future work.

The simplicity of our coplanar model prevents us from deriving any strong constraints on the orbit and mass of \PX. 
However, the dynamical mechanisms we have described serve as a starting point to inferring \PX's orbital properties and exploring consequences of the \PX~hypothesis.
We saw in Section \ref{SECN:SECULAR} and \ref{SECN:CHAOTICWEB:APSIDAL} that the mass and orbit of \PX~set the location of the stable apsidal libration island in  $q$-$\Delta\varpi$ space. 
Figure \ref{FIG:CHAOTICWEB:SECULARLARGEA} demonstrates that our fiducial choice of \PX's mass and orbit places the equilibrium at $q\sim40\au$ for distant particles, close to perihelion distances of the observational sample of clustered ESDOs. 
We propose that the loose range of \PX~masses and orbits that \BBa~and \BBb~identify as reproducing the observed clustering of ESDOs is set by the condition that these combinations produce equilibria at similar perihelion distances.

Based on our simulation results, it is tempting to conclude that \PX~should clear out all ESDOs that are not anti-aligned with its orbit beyond a critical semimajor axis $a_\text{crit}$.
However, 3D simulations show a more complicated picture at larger semimajor axes; in addition to an anti-aligned population with modest inclinations, there is a population of highly inclined orbits that exhibit different clustering behavior \citep{BatyginMorbidelli17,Li_inprep}.
Nonetheless, in 3D simulations, the secular dynamics of particles exhibits a transition near $a_\text{crit}=250\au$: interior to $a_\text{crit}$ particles remain on quasi-coplanar orbits while particles exterior to  $a_\text{crit}$ and initially nearly coplanar with \PX~are excited onto high-inclination orbits.

The mapping method developed in Section \ref{SECN:CHAOTICWEB} provides a framework for understanding the chaotic transport of distant ESDOs in the presence of \PX. 
A more complete understanding of this chaotic diffusion could, in turn, constrain the possible orbital properties of \PX.
Provided an initial distribution of TNOs, a diffusion calculation could predict the evolution of their semimajor axes under the influence of \PX~and Neptune.
The predicted loss of TNOs through this semimajor axis diffusion over the evolution of the solar system should be consistent with a reasonable initial mass of such bodies. 
Additionally, the semimajor axis distribution of observed ESDOs can be compared with the predictions of such a calculation.

\software{REBOUND, Matplotlib}

%%%%%%%%%%%%%%%%%%%%%%%%%%%%%%%%%%%%%%%%%%%%%%%%%%%%%%%%%%%%%%%%%%%%%%%%%
\section{Acknowledgments}

We thank Juliette Becker and the anonymous referee for helpful comments on draft of this manuscrpit.
M.J.H. and M.J.P. gratefully acknowledge 
NASA Origins of Solar Systems Program grant NNX13A124G, 
NASA Origins of Solar Systems Program grant NNX10AH40G via sub-award agreement 1312645088477, 
BSF Grant Number 2012384, 
NASA Solar System Observations grant NNX16AD69G, 
as well as support from the Smithsonian 2015 CGPS/Pell Grant program.
The computations in this paper were run on the Odyssey cluster supported by the FAS Science Division Research Computing Group at Harvard University.

%%%%%%%%%%%%%%%%%%%%%%%%%%%%%%%%%%%%%%%%%%%%%%%%%%%%%%%%%%%%%%%%%%%%%%%%%

\appendix

\section{Hamiltonians constructed via Numerical Averaging}
\label{SECN:APP:Equation}
Here, we describe the numerical averaging routines used to construct 
Hamiltonian contours of the secular Hamiltonian used in Section \ref{SECN:SECULAR} and the resonant Hamiltonian used in Section \ref{SECN:CHAOTICWEB:RESDYNAMICS}.
Both \citet{BEUST16} and \citet{BatyginMorbidelli17} use equivalent numerical averaging methods to study the phase space of the secular averaged problem, and in the case of \citet{BatyginMorbidelli17}, the phase space of MMRs with \PX.
We start by introducing the Hamiltonian of the elliptic restricted three-body problem in Delauney canonical variables with units chosen such that $GM_\sun = 1$. The Hamiltonian is 
\begin{eqnarray}
H_\text{three-body}(\Lambda,\lambda,L,l, G,g) = -\frac{1}{2\Lambda^2}+n_9 L - \mu_9 R\end{eqnarray}
where $\mu_9$ is the mass ratio of \PX to the Sun,
\begin{equation}
R= \left(\frac{1}{|r_9-r|} - \frac{r_9 \cdot r}{|r_9|^3} \right) \label{EQ:APP:DF}
\end{equation}
is the disturbing function, and $r$ and $r_9$ are the vector positions of the test-particle and \PX, respectively.  The canonical coordinates $(l,\lambda,g)$ are, respectively: the mean anomaly of \PX, the particle mean longitude, and the particle longitude of perihelion. Their corresponding conjugate momenta are $(L,G=\sqrt{a(1-e^2)},\Lambda=\sqrt{a})$, where $a$ and $e$ are the semimajor axis and eccentricity of the particle.\footnote{The value of $L$ is arbitrary because it does not appear in any derivatives of $H_\text{three-body}$ and therefore does not enter the equations of motion.}

\subsection{The Averaged Secular Hamiltonian}
\label{SECN:APP:SECULAR}
To construct the secular Hamiltonian used in Section \ref{SECN:SECULAR},  we also include the quadrupole contribution of Neptune, 
\begin{eqnarray}
H_\text{N,sec} = -\frac{\mu_Na_N^2}{4\Lambda^3 G^3}\,,
\end{eqnarray}
so that the full Hamiltonian is $H(\Lambda,\lambda,L,l, G,g)=H_\text{three-body}+H_\text{N,sec}$.

The secular Hamiltonian $\bar H_\text{sec}$ is computed by averaging over the orbital phases of both \PX~and the test particle and is given by the double integral
\begin{eqnarray}
\bar H_\text{sec}(G,g) &=& \frac{1}{4\pi^2} \int_0^{2\pi}\int_0^{2\pi}H(\Lambda,\lambda,L,l, G,g) dld\lambda\label{EQ:APP:HSEC} .
\end{eqnarray}
After dropping constant terms, which do not enter the equation of motion, Equation \ref{EQ:APP:HSEC} can be rewritten as 
\begin{eqnarray}
\bar H_\text{sec}(G,g) &=& -\frac{\mu_Na_N^2}{4a G^3}-\frac{\mu_9}{4\pi^2} \int_0^{2\pi}\int_0^{2\pi}\frac{(1-e\cos u)(1-e_9\cos u_9)}{|r_9-r|}dudu_9 
. \label{EQ:APP:HSEC2}
\end{eqnarray}
where $u$ and $u_9$ are the eccentric anomalies of the particle and \PX, respectively.\footnote{The change of integration variables from mean longitudes to eccentric anomalies allows the integrand to be expressed explicitly in terms of the integration variable, making numerical integration more efficient.}
We numerically evaluate the double integral in Equation \eqref{EQ:APP:HSEC2}.

\subsection{The Averaged Resonant Hamiltonian}
\label{SECN:APP:RES}
In order to study the dynamics of an $N$:$k$ resonance between a test particle and {\PX}  we introduce new canonical coordinate-momentum pairs
\begin{align}
\sigma = l-\lambda &;& \Sigma = \frac{N L +k \Lambda}{N-k} \\
 \phi = \frac{1}{k}(N \lambda - k l - (N-k)g)  &;& \Phi = \frac{\Lambda + L}{N-k} \\
\gamma = g &;& \Gamma=  G+\Lambda+L\,.
\end{align}
% \begin{eqnarray}
% k \psi = l-\lambda &;& \Psi =  -\left(j\Lambda + (j-k)L \right) \\
% k \sigma = jl -(j-k)\lambda - kg  &;& \Sigma = (\Lambda + L) \\
% g = \gamma    &;& G =  \Gamma - \Lambda-L
% \end{eqnarray}.
We can construct a model with two degrees of freedom for the resonant dynamics by averaging over $\sigma$, yielding the new Hamiltonian
\begin{eqnarray}
\bar{H}_\text{res.}(\Phi,\phi,\Gamma,\gamma) = -\frac{1}{2}\left(\Sigma-N\Phi\right)^{-2} + k n_9\Phi-\mu_9\bar{R} \label{eq:Hsyn_avg}
\end{eqnarray}
where 
\begin{eqnarray}
\bar{R} = \frac{1}{2\pi}\int_0^{2\pi} R d\sigma
\end{eqnarray}
The Hamiltonian of Equation \eqref{eq:Hsyn_avg} represents a two-degree-of-freedom system that depends parametrically on the conserved quantity $\Sigma$.
In Section \ref{SECN:CHAOTICWEB:RESDYNAMICS}, we have plotted contours of constant $\bar{H}_\text{res}$ while keeping $\Gamma$ and $\gamma=\pi$ fixed.
\section{Stroboscopic Mapping}
\label{SECN:APP:OTHER}
In this appendix, we derive the mapping used to study the dynamics of the \PX-crossing anti-aligned population in Section \ref{SECN:CHAOTICWEB}. 
The mapping approximates the change in a test particle's semimajor axis from one pericenter passage to the next to first order in \PX's mass ratio relative to the Sun, $\mu_9$. 
The change in the test particle's semimajor axis depends on its pericenter distance and longitude along with the orbital phase of \PX~at pericenter passage.
The change in the particle's semimajor axis changes its period, which in turn determines the time at which it returns to pericenter and \PX's orbital phase at this time.  

\begin{figure}
    \centering
    \includegraphics[width=\columnwidth]{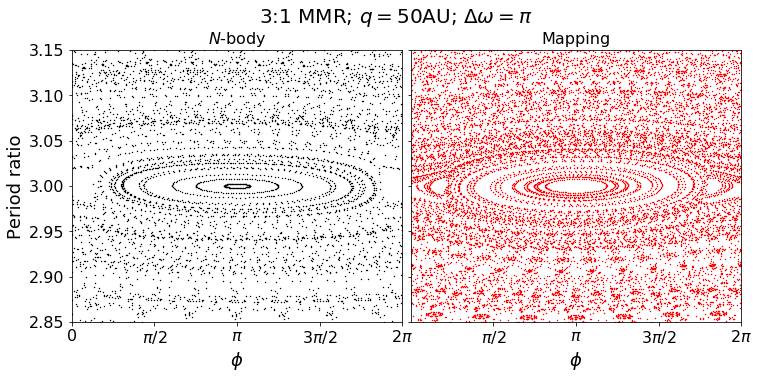}
    \caption{A comparison between the mapping model Equation \eqref{Equation:Map1} and $N$-body integration near a external 3:1 MMR with \PX.
    Trajectories are initialized at random points in the plotted range of period ratio and $\phi$ and evolved for 100 orbits. 
    To generate the $N$-body results the particle's semimajor axis (used compute the period ratio) is recorded at aphelion.
    The comparison demonstrates remarkable agreement between full $N$-body integration and our mapping approach.
    }
    \label{FIG:APPENDIXB:MAPCOMPARE}
\end{figure}

\begin{figure}
    \centering
\includegraphics[width=\columnwidth]{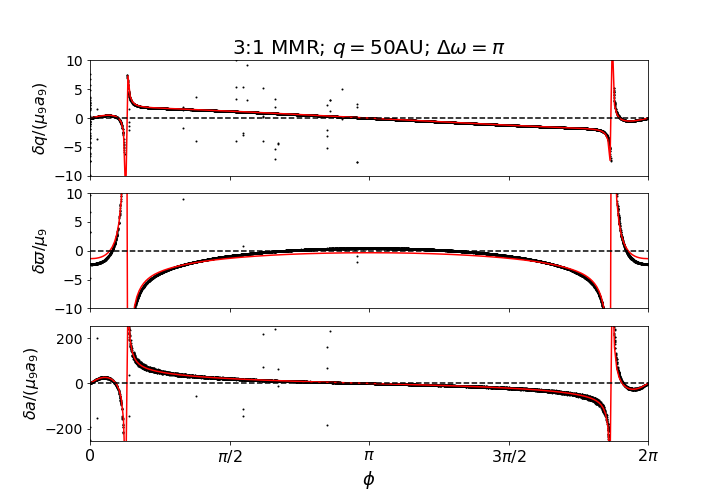}
\caption{Increments in various orbital properties after a single perihelion passage of a particle perturbed by \PX. Each panel compares analytic predictions computed with \eqref{EQ:APP:DA} shown in red against the results of $N$-body simulations, shown by black points. 
The $N$-body results are taken from the same simulations plotted in Figure \ref{FIG:APPENDIXB:MAPCOMPARE}.
The handful of scattered points that do not fall near the analytic predictions are from particles that have experienced close encounters with \PX.
}
\label{FIG:APPENDIXB:MAPCOMPARE2}
\end{figure}

To first order in \PX's mass ratio relative to the Sun, $\mu_9$, the change in any quantity, $Q(\Lambda,\Gamma,\lambda,\gamma)$ that depends on the test particle's canonical variables is 
\begin{eqnarray}
\delta Q &=& [Q,H_0]P-\mu_9 \int_{0}^{P}[Q,R] dt \label{EQ:APP:DA}
\end{eqnarray}
where $P$ is the test particle's orbital period, $[\cdot,\cdot]$ is the Poisson bracket, $R$ is the disturbing function given by Equation \eqref{EQ:APP:DF}, and the integral is evaluated along the unperturbed Keplerian orbit of the test particle.
To construct the mapping used in Section \ref{SECN:CHAOTICWEB}, we evaluate the integral in Equation \eqref{EQ:APP:DA} by approximating the test-particle trajectory as a parabolic orbit that passes through pericenter at time $t=0$.
We  extend the integration limits from $\pm\infty$ and use the trajectory of a parabolic orbit in the integral. 
To compute increments in various orbital quantities using Equation \ref{EQ:APP:DA},
we must  express the trajectory of the particle orbit orbit in terms of canonical coordinates.
Following \cite{PetroskyBroucke88}, we use the canonical pair $(E,t)$, the orbital energy and time (relative to perihelion passage), rather than $(\Lambda,\lambda)$, because the former are well-defined for both bound and unbound orbits.\footnote{The canonical transformation between the coordinate pairs can be generated with the type-3 generating function $F_3=-\Lambda^{-2} t/2$.}
The parabolic trajectory of a particle, in canonical coordinates, is given by
\begin{eqnarray}
\vec{r}&=&\frac{G^2}{2}\left\{(1-\tau^2)\cos g -2\tau\sin g,2\tau\cos g + (1-\tau^2) \sin g \right\}\nonumber\\
t &=&  \frac{1}{4}G^3\left(\tau+\frac{1}{3}\tau^3 \right)~.
\end{eqnarray}
In Figures \ref{FIG:APPENDIXB:MAPCOMPARE} and \ref{FIG:APPENDIXB:MAPCOMPARE2}, we compare $N$-body results to our mapping, Equation \eqref{Equation:Map1}, and analytic predictions for $\delta q,\delta \varpi$, and $\delta a$ using Equation \eqref{EQ:APP:DA}\footnote{When computing the integral in Equation \eqref{EQ:APP:DA} for a parabolic test-particle trajectory,
terms involving the indirect term of the disturbing function, $|r_9|^{-3}(r_9 \cdot r)$, diverge. 
We replace the indirect term instead with $|r|^{-3}(r_9 \cdot r)$. It is straightforward to show that this replacement is equivalent, up to order $\mu$, to a transformation of the Hamiltonian in Equation \eqref{EQ:APP:DF} from heliocentric canonical coordinates to barycentric canonical coordinates.}, demonstrating remarkable agreement between full $N$-body integration and our mapping approach.

The mappings plotted in Figure \ref{FIG:APPENDIXB:MAPCOMPARE2} diverge at the two $\phi$ values that lead to an immediate collision with \PX.
Approximating the motion to first-order in $\mu$ is no longer valid very near these discontinuities, and a handful of points that have experienced close encounters appear scattered away from the analytic predictions in Figure \ref{FIG:APPENDIXB:MAPCOMPARE2}.
To implement our mapping in Equation \eqref{Equation:Map1}, we  interpolate the function $\delta E$ at a finite number of $\phi$ values, so these discontinuities become somewhat smoothed out.
%%%%%%%%%%%%%%%%%%%%%%%%%%%%%%%%%%%%%%%%%%%%%%%%%%%%%%%%%%%%%%%%%%%%%%%%%
\newpage 
\bibliography{references}

\newpage 
\listofchanges

%%%%%%%%%%%%%%%%%%%%%%%%%%%%%%%%%%%%%%%%%%%%%%%%%%%%%%%%%%%%%%%%%%%%%%%%%

\end{document}